\documentclass[]{aastex631}

\usepackage{multirow}
\usepackage{amsmath}
\usepackage{hhline}
\usepackage{array, makecell}
\usepackage{booktabs, tabularx}
\usepackage{comment}
\usepackage{rotating}
\usepackage{enumitem}
\usepackage{mathrsfs}

\usepackage{float}
\usepackage{xcolor}
\newcounter{tr}
\ifnum \value{tr}>5

\newcommand{\deletedD}[1]{{\color{red} Damien - Deleted: } \sout{#1}}

\newcommand{\authorcommentD}[1]{{\color{blue} Damien - Comment :} {\color{blue} #1}}

\else

\newcommand{\deletedD}[1]{}

\newcommand{\authorcommentD}[1]{}

\fi

\begin{document}


\title{X-ray Analysis of Gamma-Ray Burst Flares and Underlying Afterglows: \\ Insights into Origin of Flares}

\shorttitle{Flares vs prompt and afterglow }


\shortauthors{Dereli-B\'egu\'e et al.}

\author[0000-0002-8852-7530]{H. Dereli-B\'egu\'e}
\affiliation{Department of Physics, Bar-Ilan University, Ramat-Gan 52900, Israel}

\author[0000-0001-8667-0889]{A. Pe{'}er}
\affiliation{Department of Physics, Bar-Ilan University, Ramat-Gan 52900, Israel}

\author[0000-0003-4477-1846]{D. B\'egu\'e}
\affiliation{Department of Physics, Bar-Ilan University, Ramat-Gan 52900, Israel}

\author[0000-0002-9769-8016]{F. Ryde}
\affiliation{Department of Physics, KTH Royal Institute of Technology and The Oskar Klein Centre, SE-106 91 Stockholm, Sweden}

\author[0009-0006-7782-0164]{A Gowri}
\affiliation{Indian Institute of Science Education and Research (IISER), Mohali, India}
\affiliation{Department of Physics, Bar-Ilan University, Ramat-Gan 52900, Israel}


\begin{abstract}
Gamma-ray burst (GRB) X-ray light curves exhibit a variety of complex temporal structures, such as flares and plateaus. The origin of flares seen in many GRB early afterglows is still uncertain. Here, we analyze a sample of 89 GRBs, 61 of them with flares, both with and without a ``plateau" phase. We fit the Swift-XRT light curves with synchrotron emission from a forward shock propagating into either a constant-density ISM or a stellar wind, and flares on top of that. We find that the flare light curves are not symmetric, with a decay time that is $\sim$five times longer than the rise time. We do not find any differences in flare properties between GRBs with and without a ``plateau" phase. Moreover, additional afterglow properties such as the electron power-law index and the end time of the plateau are consistent between bursts with and without flares. These results strongly indicate that flares originate from a mechanism distinct from that producing the plateau and afterglow. When looking at the prompt emission properties, we do find some tendencies: GRBs with flares tend to be brighter and longer lasting than GRBs without flares. We therefore conclude that, unlike plateaus, flares are unlikely to arise from an external origin and are more plausibly associated with prolonged central engine activity that lasts longer than the main episode that produces the prompt phase. As the plateau cannot have the same origin, this result excludes models of late-time energy injection as the source of the GRB plateau.
\end{abstract}

\keywords{Gamma-ray bursts, Light Curves: X-ray, Astronomy data analysis, Relativistic jets, Radiation mechanisms: non-thermal.}


\section{Introduction} 
\label{sec:intro}

Gamma-ray bursts (GRBs) are among the most luminous explosions in the universe, typically consisting of two emission phases: the prompt $\gamma$-ray burst and the longer-lasting afterglow. The classical (standard) fireball model \citep{RM92, MRW98, Pir04, Mes06, KZ15} attributes the prompt emission to internal energy dissipation within a relativistic jet and the afterglow to synchrotron radiation from an external shock propagating into the surrounding medium. This circumburst environment can be either a stellar wind or a constant-density interstellar medium (ISM) \citep{MR97, SPN98}.

Following the launch of the Neil Gehrels Swift Observatory (hereafter Swift; \citealt{Gehrels+04}), detailed observations of the X-ray band during the afterglow phase revealed several unexpected features in GRB light curves that were not predicted by the standard model \citep{Nousek06, OBrien06, Zhang06, Evans2007, Evans09}. These include early steep decays, plateau phases, and particularly X-ray flares, characterized by sudden, high-amplitude flux variations. Despite being observed for 20 years now, there is still no consensus on the origin of these phenomena and connection, if any.

The late-time X-ray decay ($F_x (t) \propto t^{-\alpha}$ with $1.2<\alpha<1.5$ before a jet break) aligns with predictions of the standard model. As opposed to that, 
the early steep decay and plateau phases, which are key components of GRB X-ray afterglows, were not predicted. The steep decay phase, typically characterized by a temporal index of $3 \leq \alpha \leq 5$, is interpreted as high-latitude emission from the tail end of the prompt phase \citep{BCG05, TGC+05, WOO07, ROB+21}. It is often followed by a plateau phase with a shallower decay slope ($0 \leq \alpha \leq 0.7$), which generally occurs between $10^2$ and $10^3$ seconds after the GRB trigger and transitions into a steeper decay phase, consistent with external shock expectations \citep{Zhang06, DPR22}. 

The plateau phase, observed in nearly 60\% of GRBs \citep[][and references therein]{Evans09, Dainotti+2010, Dainotti+2017, Srinivasaragavan20}, has been linked to prolonged central engine activity, including energy injection by a magnetar or accreting black hole \citep{Nousek06, Metzger11, DallOsso+2011, Cannizzo+2011, vanEerten2014a, vanEerten2014b, Ronchini+2023, Lenart+2025}. Alternative explanations involve structured jets seen off-axis \citep{EG06, Eichler14, BDD+20a}, or low Lorentz-factor outflows in a wind environment \citep{ShM12, DPR22}. This last interpretation was strengthen by our recent analysis presented in \citet{DB+25}. In that work, we separated the sample into GRBs with and without plateau phases, and analyzed the properties of X-ray flares in both samples. We found that the distributions of flare characteristics, such as peak time and the width-to-peak-time ratio, are remarkably similar across both groups. This finding lends support to the scenario proposed by \citet{DPR22}, in which GRBs with X-ray plateaus are characterized by lower Lorentz factors, as the resulting flares in these bursts are produced at smaller radii but with similar temporal properties. This is in contrast to the predictions of the alternative models, in which one would expect a statistical difference between the properties, such as delay time of the flares between GRBs with and without a ``plateau" phase. 

Indeed, the origin of X-ray flares and their relation to other GRB properties are  still open questions. These flares are typically observed between $100$ and $10^5$ seconds after the prompt emission. They are observed in about half of the GRB population, mostly in long GRBs, and only rarely in short ones \citep{Burrows+05, Falcone+06, Falcone+07, Chincarini+07, Curran+08, Chincarini+10, MBB11, Bernardini+11}. They usually appear as one or two flares, with cases of multiple flares being rare \citep[e.g.,][]{Perri+07, Abdo+11}.

A fundamental question remains regarding the origin of these flares. Flare have been extensively compared to the prompt emission, since both exhibit several common properties, such as: (i) hard spectra \citep{Burrows+05, Romano+06, Falcone+06};  (ii) hard-to-soft evolution \citep{Falcone+07}; (iii) similar intensity ratios between successive emission episodes \citep{Chincarini+07}; (iv) a positive correlation between spectral lag and luminosity \citep{Margutti+10}; (v) a significant positive correlation between pulse rise times and the minimum variability timescale \citep{Sonbas+13}; (vi) a power-law waiting time distribution shared by prompt pulses and flares \citep{Guidorzi+15}. Here, the waiting time is defined as the time interval between successive emission episodes.  Its similar behavior in both prompt emission and flaring activity, suggests that a common underlying physical process may be responsible for both.

These similarities support the widely accepted view that flares are powered by late-time central engine activity, either via internal shocks \citep{IKZ05, FW05, Zhang06} or other dissipation processes within the ultra-relativistic outflow \citep{Giannios06, Lazzati+11}, slower jet \citep{BK16} or structured jet \citep{DBD+22} but occurring at later times and with lower energies. However, \citet{SIM23} recently showed a trend of positive correlation between the minimum variability timescale of flares and their duration, and of strong negative correlation with peak flux not seen in the prompt emission. Additionally, the strong positive correlation observed between flare width and peak time is not seen in the prompt emission properties \citep{Ramirez-Ruiz+2000, Chincarini+07, Chincarini+10, DB+25}.
To explain such observational differences, several alternative models have been proposed, all involving accretion onto a black hole. These include fragmentation of a rapidly rotating stellar core \citep{King+2005}, magnetic regulation of the accretion flow \citep{Proga&Zhang2006}, and fragmentation in the outer part of the accretion disk caused by gravitational instability \citep{Perna+2006}, instability in the fall-back material \citep{Kumar+2008a}. In this work, we interpret our observational results within the framework of the accretion models.

Furthermore, only a few studies have compared the flare properties to those of the underlying afterglow. These show that the afterglow typically maintains a consistent temporal slope before and after the flare, which most likely rules out an external shock origin \citep{Burrows+06, Chincarini+06, Chincarini+07}.
Other work with a few GRBs have shown that the late-time decay slope does not correlate with the flare properties \citep{Curran+08}. And another study discuses a correlation between early-time luminosity
at 200 s and the average temporal decay onwards after that time and show that the correlation is unaffected by the presence of flares or plateaus \citep{Racusin+16}.

In our previous paper \citep{DB+25}, we considered a possible connection between the properties of the flares and the existence of a plateau, in order to constrain theoretical models on the origin of the plateau phase. We showed that the flares peak time and the ratio of the flare width to the flare peak time are similar at bursts with and without a plateau, and we therefore deduced that both are unlikely to share a common origin, such as late-time energy injection. Here, we further extend this work, to study additional flare properties that were not previously considered, to put stronger constraints on the underlying mechanism of flare production and its relation (or lack thereof) to the underlying mechanisms responsible for the prompt, plateau and afterglow phases in GRB evolution. We therefore present a comprehensive re-analysis of several GRBs observed with Swift-XRT and discuss the results of data obtained by both Swift-XRT and BAT as well as Fermi-GBM for a few GRBs. 

This paper is organized as follows. In Section \ref{sec:sampling_modelling_fitting}, we summarize the sampling, modeling and fitting procedure. In Section \ref{sec:afterglow_analysis_results}, we present the afterglow analysis results for four subsamples (with/without flares and with/without plateaus), dividing the results into subsections on statistical analysis, flare properties, and underlying afterglow characteristics. In Section \ref{sec:prompt_properties}, we compare the prompt properties for these four subsamples. In Section \ref{sec:discussion} and \ref{sec:asymmetry_implication}, we then discuss the physical implication of the results for the origin and and nature of GRB flares and plateaus.
Finally, in Section \ref{sec:conclusion}, we list our summary and conclusions.

\section{Data and Modeling Approach}
\label{sec:sampling_modelling_fitting}

We use the same sample and the same fitting procedure as published earlier \citep{DB+25}. We therefore provide here a brief summary, and refer the reader to \citet{DB+25} for a detailed description. 

\subsection{Sample Selection}

We constructed a statistically significant sample of 100 GRBs by searching the Swift archive in reverse chronological order, starting from 2022 December 2, and covering the previous 8 years. The selection procedure builds directly on our earlier study \citep{DPR22}, using the same robust criteria: (i) a known redshift \footnote{\url{https://www.mpe.mpg.de/~jcg/grbgen.html}}, which is essential for calculating energetics; (ii) GRBs which triggered the Swift Burst Alert Telescope (Swift-BAT) detector, since GRBs not detected by BAT usually lack  Swift X-Ray Telescope (Swift-XRT) data necessary for our light curve analysis and flare fitting procedures. (iii) We further excluded GRBs with insufficent number of data points, by requiring at least 5 data points at the beginning and end of each independent segment: plateau or late-time afterglow decay.   
 These segments were visually inspected and identified based on the light curves available through the Swift online repository.

\subsection{Light curve modeling} 
\label{sec:model}
X-ray afterglow light curves typically exhibit multiple temporal segments: (i) an initial steep decay, (ii) a plateau phase (if present), (iii) a late-time afterglow power-law decay (interpreted as the self-similar deceleration of the forward shock), and (iv) a post-jet break phase. These are occasionally interrupted or superimposed with (v) X-ray flares. When all components are present, they form the so-called ``canonical" X-ray afterglow \citep{Nousek06, OBrien06, Zhang06}. However, the diversity of light curves across GRBs often presents only a subset of these components, complicating model selection and interpretation.

To model this diversity, we adopted a modular fitting strategy using 36 candidate models composed of one to four connected power-law segments, optionally including up to two flares and a jet break. Flares were modeled with the asymmetric ``Norris" function \citep{NBK05}, following the full formulation and parameterization outlined in \citet{DB+25}, which has been shown to provide a superior representation of GRB flare profiles compared to symmetric or phenomenological alternatives as discussed in several papers \citep[e.g.,][]{Chincarini+07, Bernardini+11}. This approach allows for the determination of key flare properties, including amplitude, peak time, temporal width $ w $, flare rise and decay times and asymmetry $k$. The width was defined in log-space based on half-maximum limits, and the flare energy $E_{\rm iso,f}$ was computed via numerical integration. Although width definitions can vary across studies, the ratio $w / t_{\rm pk}$ provides a robust and consistent descriptor of flare duration and variability.
The complete set of equations describing key flare parameters such as the peak time $t_{\rm pk}$ and temporal width $w$ are presented in \citet{DB+25}. For completeness, here we provide the equations for the flare rise and decay times:
\begin{equation}
    t_{\rm rise} = t_{\rm pk} - 10^{\bar t_1}, 
    \quad
    t_{\rm decay} = 10^{\bar t_2} - t_{\rm pk},
    \label{eq:trise_tdecay}
\end{equation}
where $\bar{t}_1$ and $\bar{t}_2 (> \bar{t}_1)$ are the times at which the Norris function reaches  half of its maximal value\footnote{We denote these times as $\bar{t}$ to
avoid confusion with the afterglow break times $T_1$, $T_2$, and $T_3$.}. A key methodological difference from previous works is that our fits are performed in log-space. 
This differs from the definitions adopted in earlier studies \citep[e.g.,][]{NBK05, Chincarini+10}, which relied on linear-space fitting.

Our afterglow modeling relies on physically motivated assumptions: the afterglow is produced by synchrotron emission from shock-accelerated electrons in a relativistic blast wave, propagating either into a constant-density interstellar medium (ISM) or a stellar wind profile. As shown by several studies \cite[e.g.,][]{GS02, DPR22}, this framework leads to characteristic light curve behaviors based on the location of the observed frequency relative to the synchrotron spectral breaks $\nu_{\rm m}$ and $\nu_{\rm c}$, the power-law index $p$ of the electron energy distribution, and the dynamical phase of the outflow.

The inclusion of a steep decay segment, jet break, and up to two flares allows for capturing the full complexity of Swift-XRT observations while maintaining a balance between physical interpretability and statistical flexibility. Although our models assume spherical symmetry, we note that deviations from this assumption primarily affect the post-jet break behavior, whereas flares are observed at earlier times and are therefore largely unaffected. Combined together, we consider a total of 36 different models used in fitting the data. These 36 models are constructed as follows. First, we separate based on density profiles: ``wind", vs the ``constant-density ISM". A ``plateau" can only be produced by models in the "wind" category. This category is divided into 12 models, differ by (i) the presence or absence of a jet break, (ii) the number of flares (0, 1, or 2), and (iii) the synchrotron spectral regime. All 12 models assume the presence of an early steep decay.  
The ``ISM" scenario allows for both the presence or absence of the steep decay, in addition to the above variations, resulting in total of 24 models. A complete description of the models is presented in \citet{DB+25}.

\subsection{Fitting Procedure} 
\label{sec:Fitting_procedure}

The light curves were retrieved from the online Swift-XRT repository\footnote{\url{https://www.swift.ac.uk/xrt_curves/}} \citep{Evans2007, Evans09}, using the full (0.3–10 keV) bandpass. Each light curve was independently fit with all 36 different model configurations.  
To avoid introducing subjective bias, we deliberately chose not to pre-classify light curves into categories based on their morphology. Instead, our approach allows for a fully automated model comparison across the all considered models. This enables an unbiased evaluation of whether a flare or plateau is statistically and physically required by the data, ensuring consistency with our previous work \citep{DPR22}. Model selection was performed using multiple statistical criteria, including the Akaike information criterion (AIC), the corrected AIC (AICc), and Bayes Factors, as described in \citet{DB+25}. 

Once the fits were completed and the analysis performed, we excluded 11 GRBs from the remaining analysis due to misidentified flares, which resulted in the failure of different afterglow models to adequately fit the data. Our final sample consists of 89 (out of 100) GRBs,
without any separation based on the existence or absence of flares or plateaus and presented in Appendix Table~\ref{tab:fit_params1}.


\section{Afterglow analysis results}
\label{sec:afterglow_analysis_results}

\subsection{Statistical results}
\label{sec:statistical_results}
A statistical analysis of the 89 remaining bursts in the sample show the following characteristics:
\begin{enumerate}
\item Approximately $69\%$ of all GRBs (61 GRBs) have flares, indicating that flares are quite
common in GRB afterglows. This value is slightly higher than previously reported, where about half of the GRBs were found to exhibit flares \citep{Zhang06, Chincarini+10, Racusin+16}. 
\item About $67\%$ of GRBs with flares have a plateau,
\item About $73\%$ of GRBs with a plateau also exhibit flares.

Interestingly, the overall occurrence of flares in the entire sample does not differ from the occurrence
of flares in the GRB subsample with a plateau phase. This suggests that the presence of flares is
independent of the existence of a plateau, indicating that these two phenomena, namely plateaus and flares,
might not be related or dependent on each other. 

\item Moreover, we find that about $80\%$ of all GRBs in our sample exhibit steep decay phases, while a jet break is observed in $36\%$ of them, in agreement with previous studies \citep[e.g.,][]{Zhang06}. 
\end{enumerate}

To investigate the underlying afterglow properties (Subsection \ref{sec:afterglow_properties}) and prompt emission properties (Section \ref{sec:prompt_properties}), we divide the full sample into four groups. Among the 61 GRBs that exhibit flares, 42 show a plateau phase and 19 do not. Similarly, out of the 28 GRBs without flares, 15 display a plateau and 13 do not.

\subsection{Flare properties}
\label{sec:flare_properties}
\subsubsection{GRBs with flares: with and without Plateaus}
\label{sec:flare_properties1}
In \citet{DB+25}, we only study the flare properties when dividing the sample of 61 GRBs that show flares into two subsamples: those with a plateau phase (42 GRBs/65 flares) and without it (19 GRBs/32 flares). We focused on the flare peak time and the ratio of flare width to peak time. We found that the temporal properties of flares, including their peak times and the ratio of their width to peak time ($w/t_{\rm pk}$, which is found to be approximately one regardless of the presence of a plateau), remain statistically consistent across both subsamples, indicating that the presence or absence of a plateau phase does not influence the observed characteristics of flares. Here, we study additional flare characteristics, including rise time, decay time, asymmetry, and correlations between these properties.

\subsubsection{Relations between the flare parameters} 
\label{sec:correlations2}
One of the properties characterizing the temporal behavior of a flare is its asymmetry. There are two ways of displaying asymmetry in flares, either by studying the flare asymmetry constant $k$ as is presented in \citet{DB+25} or by analyzing the ratio between the rise time 
$t_{\rm rise}$ and decay time $t_{\rm decay}$ (see equation \ref{eq:trise_tdecay} in Section \ref{sec:model}). In this work, we focus on the second option, as we further aim to examine the relationship between the rise and decay times, whose values are provided in Appendix Table~\ref{tab:fit_params1}. 

\begin{figure}
\centering
\includegraphics[width=0.45\linewidth]{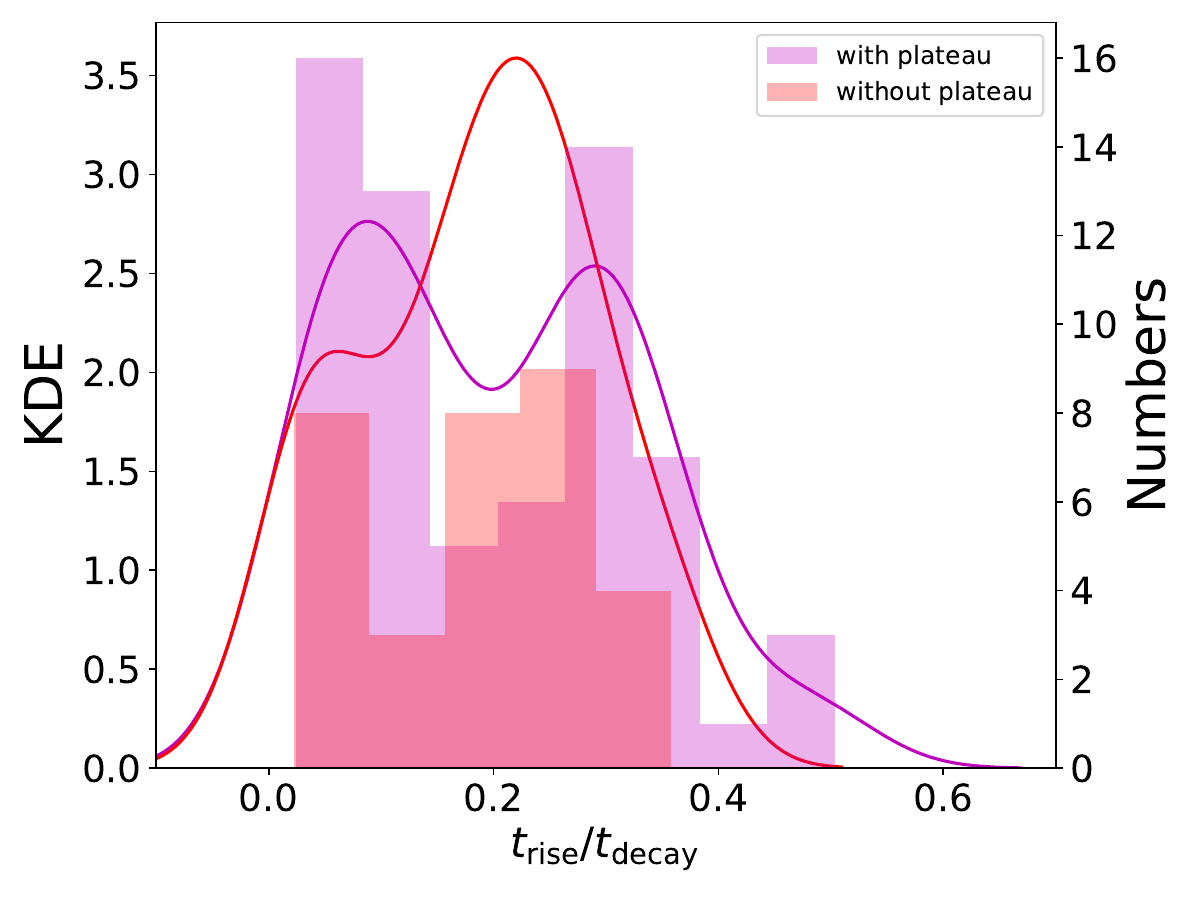}
\includegraphics[width=0.45\linewidth]
{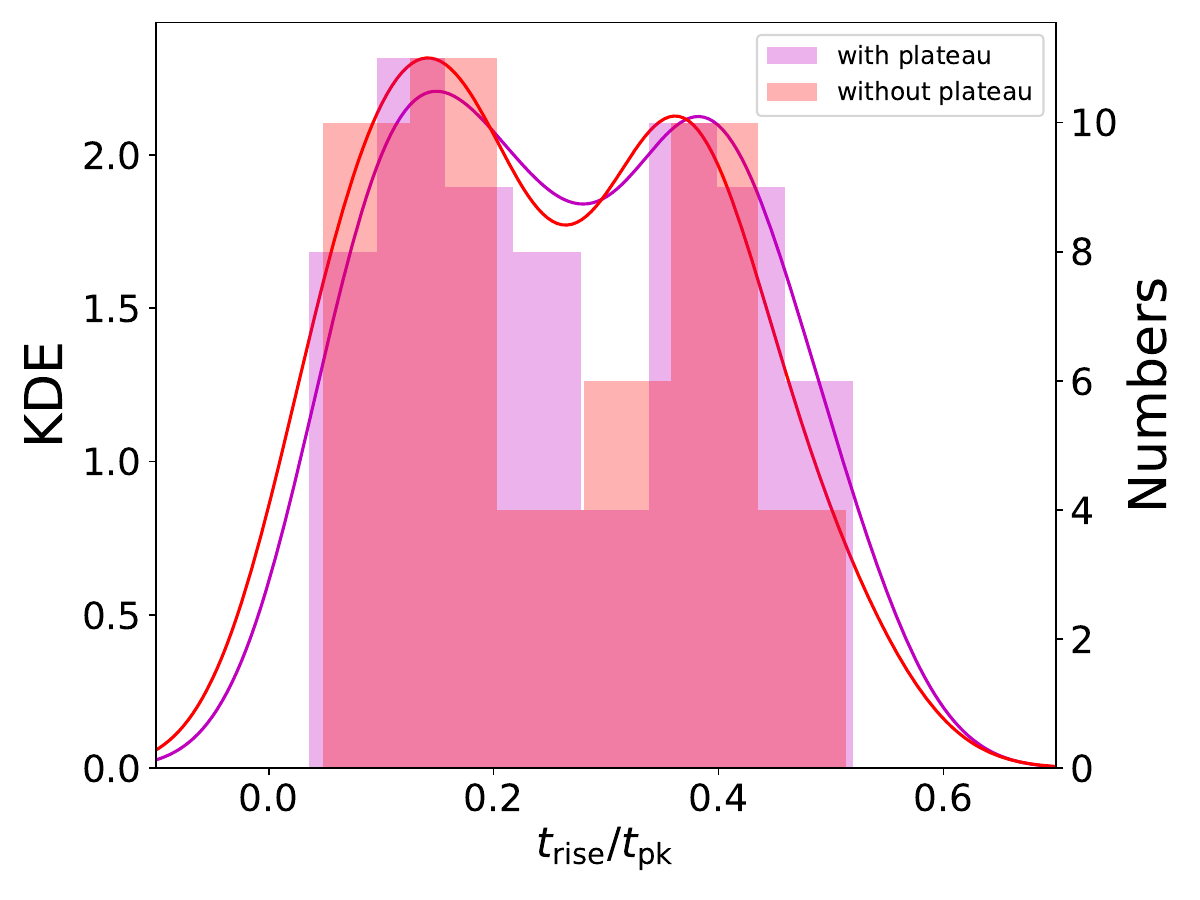}
\caption{Left: Distributions of the flare asymmetry, $t_{\rm rise}/t_{\rm decay}$ ratio. Right: Distributions of $t_{\rm rise}/t_{\rm pk}$ ratio. The purple bars represent the 42 GRBs (65 flares) with a plateau,
while the red bars represent the 19 GRBs (32 flares) without a plateau. In each panel, the right-hand ordinate shows the number of bursts in each bin while the left-hand ordinate shows the value of the kernel density estimation (KDE) drawn by the purple and red solid lines for each sample respectively.
The results show that flares exhibit a pronounced asymmetry, regardless of whether a plateau phase is present in the GRB X-ray light curves. We also observe a bimodality in both distributions of $t_{\rm rise}/t_{\rm decay}$ ratio, which is more pronounced for GRBs with a plateau phase. A stronger bimodality is equally evident in both subsamples for the $t_{\rm rise}/t_{\rm pk}$ ratio.}
\label{fig:flare-rise_decay_ratio}
\end{figure}

In the left panel of Figure~\ref{fig:flare-rise_decay_ratio}, we present the ratios 
$t_{\rm rise}/t_{\rm decay}$ and find that the mean values for the subsamples with and without a plateau 
are 0.20 and 0.19, respectively, each with a standard deviation of 0.02. 
These values are significantly lower than those reported by \citet{Chincarini+10} 
for flares observed in \textit{Swift} GRBs 
($\langle t_{\rm rise}/t_{\rm decay} \rangle = 0.49 \pm 0.26$) 
and by \citet{KRL03} for separable BATSE prompt pulses 
($\langle t_{\rm rise}/t_{\rm decay} \rangle = 0.47$), 
indicating a systematically stronger asymmetry in our sample. 

The difference between our results and those of \citet{Chincarini+10} arises primarily 
from the fitting method used in our analysis. 
First, to ensure robust flare identification, 
we applied a flux threshold to prevent small random fluctuations in the light curve 
from being misidentified as flares 
(see \citealt{DB+25}, Sec.~4). 
Second, we fit the data in logarithmic space rather than linear space. Third, 
we physically modeled the afterglow emission as synchrotron radiation from 
shock-accelerated electrons in a spherical forward shock. 
This model accounts for different blast-wave evolution stages and external medium profiles, 
with flares superimposed on the underlying continuum. 
These modeling choices and their advantages are discussed in detail in \citet{DB+25}. 

We further performed a K-S test for the distributions shown in the left panel 
of Figure~\ref{fig:flare-rise_decay_ratio}. 
The test yields a K-S statistic of $D = 0.21$ with a probability $p = 0.23$, 
indicating that the distributions for GRBs with and without plateaus are 
statistically consistent with being drawn from the same population. 
Therefore, the presence of a plateau does not have any impact on the strong 
flare asymmetry observed in our sample.

In the right panel of Figure~\ref{fig:flare-rise_decay_ratio}, we present the ratios $t_{\rm rise}/t_{\rm pk}$ and find that the mean values of the ratios $t_{\rm rise}/t_{\rm pk}$ are 0.27 and 0.25 for both subsamples (with and without a plateau), each with a standard deviation of 0.02.  We further performed  a K-S test for both subsamples and found that the test statistic is $D = 0.14$ with a probability $p = 0.74$ which show that both distributions are drown from the same population. These results show that $t_{\rm rise}/t_{\rm pk}$ exceeds the commonly assumed dissipation limit ($\delta t/t < 0.1$), which is typically required for a purely internal origin and used to rule out the possibility that flares are produced as a result of inhomogeneity in the circumstellar medium \citep{Nakar&Granot2007}. Interpretation of this result is provided in the discussion Section~\ref{sec:asymmetry_implication}.

\begin{figure}
\centering
\includegraphics[width=0.45\linewidth]{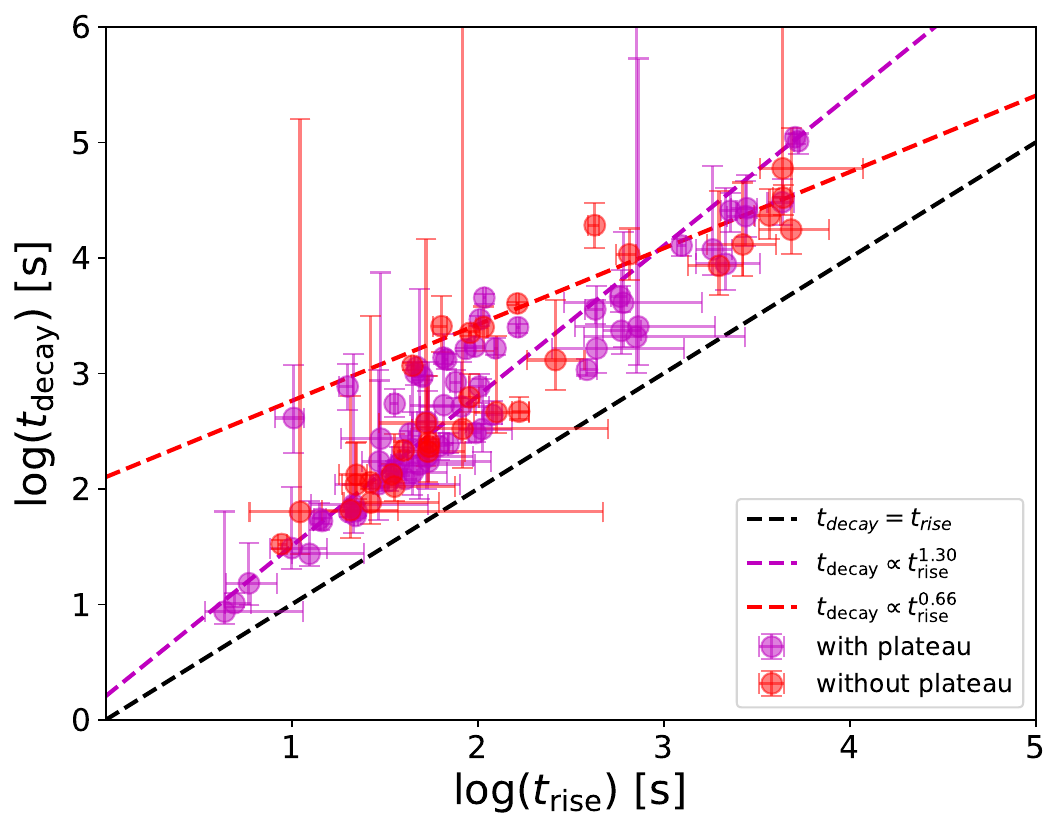}
\caption{Relation between the flare rise time ($t_{rise}$) and the flare decay time ($t_{decay}$). 
The purple points represent the 42 GRBs (65 flares) with plateau phases, the red points
represent the 19 GRBs (32 flares) without plateau phases in our subsamples. The errors correspond to a significance of one sigma. The Spearman’s rank correlation coefficient is $r = 0.90~(0.94)$,
with a corresponding chance probability of $p \ll 10^{-5} (\ll 10^{-5})$ for GRBs with (without) a plateau phase. 
This strong correlation between the flare decay and rise times is not changed when separating the first and second flares.}
\label{fig:scatter_trise_tdecay}
\end{figure}

The relationship between $t_{\rm rise}$ and $t_{\rm decay}$ is presented in Figure \ref{fig:scatter_trise_tdecay}. A clear correlation is observed for both subsamples, with (without) plateau. The result of the Spearman’s rank correlation coefficient is $r = 0.90~(0.94)$ with a corresponding chance probability of $p \ll 10^{-5} (\ll 10^{-5})$,
showing a strong positive correlation between $t_{\rm rise}$ and $t_{\rm decay}$ for both subsamples with and without plateau respectively. We then performed a fit to the correlation using the functional form
$t_{\rm decay} = C \cdot t_{\rm rise}^r$, where $C$ is a proportionality constant and $r$ is a power-law index. To account for uncertainties, errors were capped at the 90$^{th}$ percentile of the reduced distribution for the sample with a plateau, and at the median value for the sample without a plateau. We found the power-law index
$r = 1.3 \pm 1.6 \times 10^{-6}$ for bursts with a plateau and  $r = 0.66 \pm 1.2 \times 10^{-6}$ for bursts without a plateau.
The correlation hold even when separating the first and second flares, as we have checked.
These correlations are consistent with the correlation
found by \cite{Chincarini+10} ($0.79 \pm \ll 10^{-5}$) for GRBs with flares.

In conclusion, flares are asymmetric, with decay times \( t_{\rm decay} \) averaging about five times longer than the rise times \( t_{\rm rise} \), in agreement with, but quantitatively stronger than, the findings of \citet{Chincarini+10} where they found that the decay times $t_{\rm decay}$
averaging twice the rise times $t_{\rm rise}$. We find very strong correlations between flare rise and decay times for both subsamples. We find no significant difference in this behavior between the subsamples with and without a plateau phase, a result that, to our knowledge, has not been reported before. Moreover, separating the first and second flares did not change the result.

\subsection{Underlying afterglow properties}
\label{sec:afterglow_properties}

\subsubsection{Properties of subsamples with and without plateaus and flares}
\begin{figure}[H]
    \centering
    \includegraphics[width=0.45\linewidth]{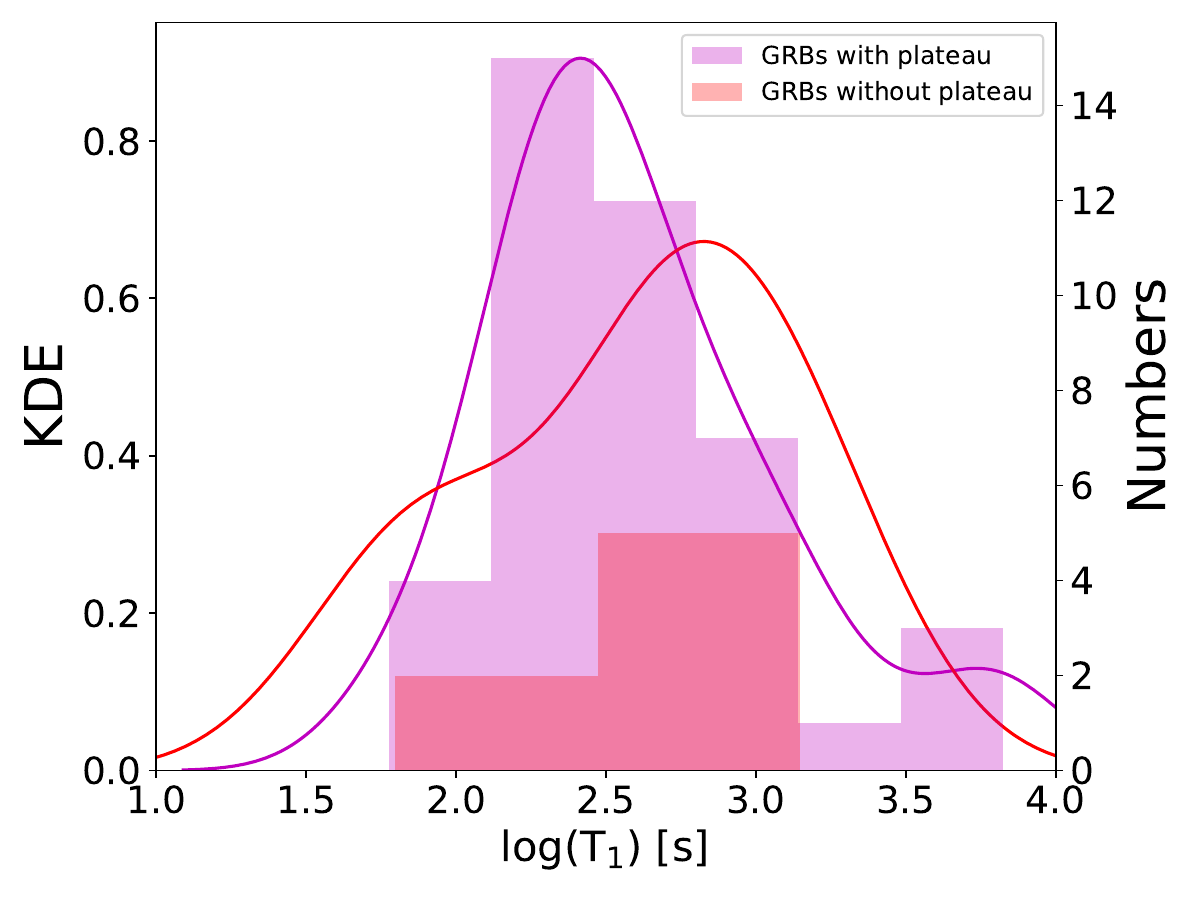}
    \includegraphics[width=0.45\linewidth]{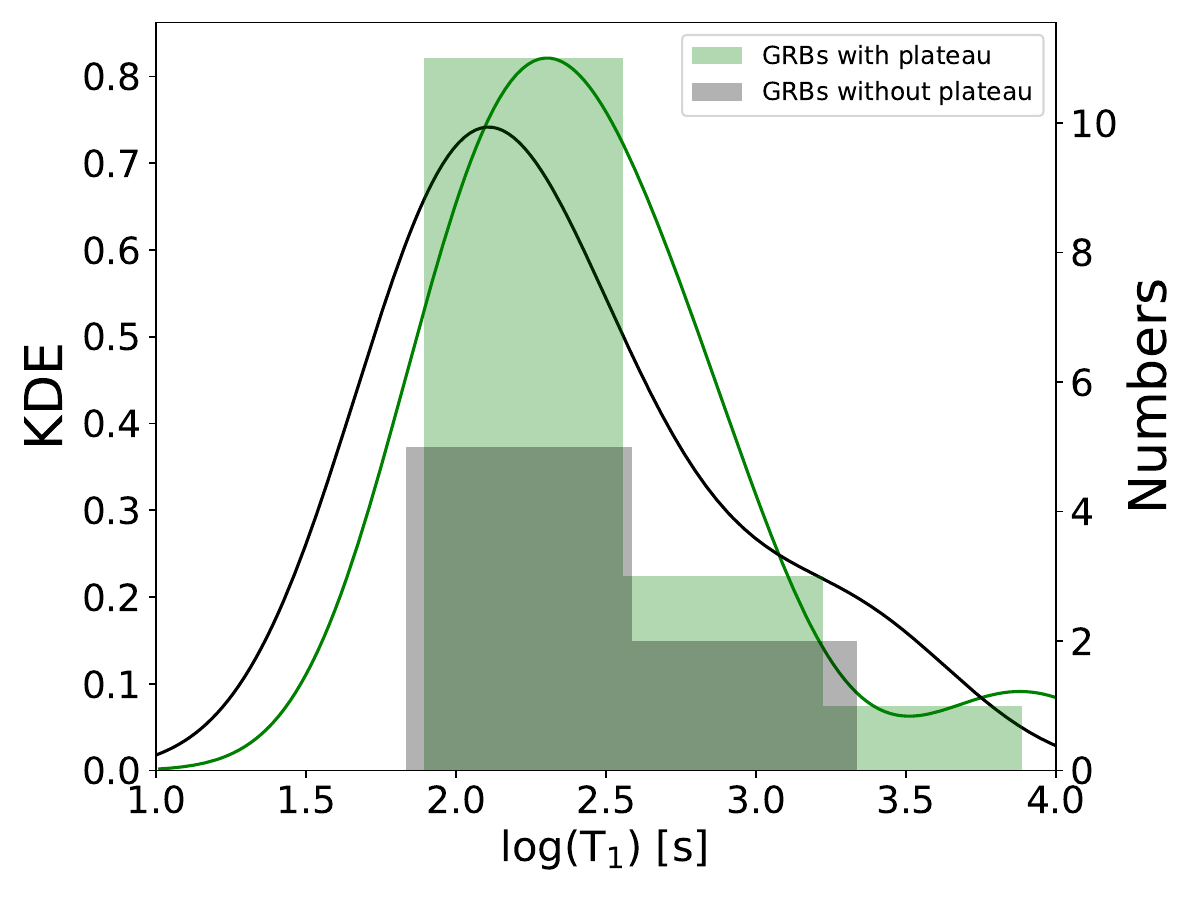}
    \caption{Distributions of the end time of the steep decay, $T_1$, which marks the beginning of the first afterglow segment (see Appendix A in \citet{DB+25}) for GRBs with flares (left) and without flares (right), with each panel showing the subsamples with and without plateaus. In the left panel, the 42 GRBs with a plateau phase are shown in purple, and the 19 GRBs without a plateau phase in red. In the right panel, the 15 GRBs with a plateau phase are shown in black, and the 13 GRBs without a plateau in green. 
    In each panel, the right-hand ordinate shows the number of bursts in each bin while the left-hand ordinate shows the value of the kernel density estimation (KDE) drawn by solid lines corresponding to each color. These distributions show no significant difference at $T_1$ across the subsamples.}
    \label{fig:time_after_early_steep_decay}
\end{figure}

\begin{figure}[H]
    \centering
    \includegraphics[width=0.45\linewidth]{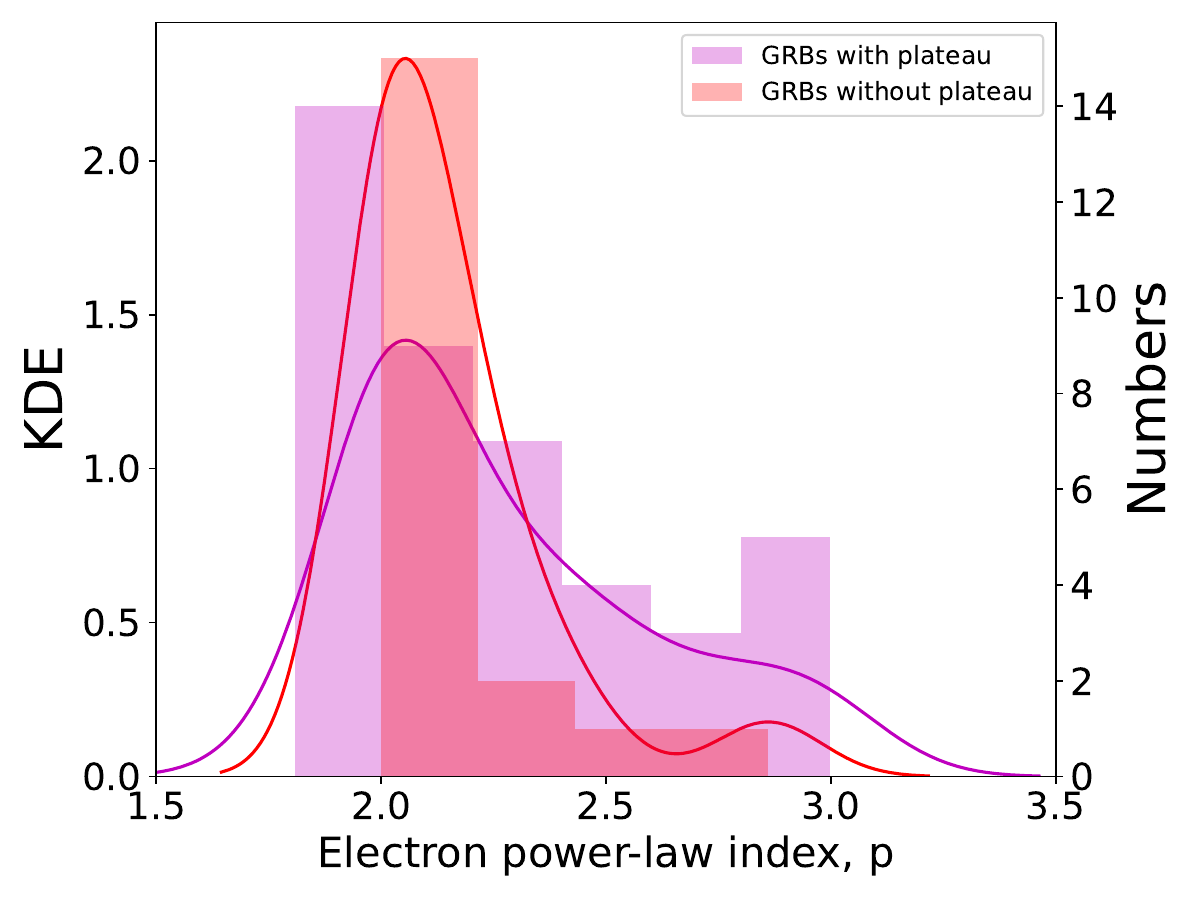}
    \includegraphics[width=0.45\linewidth]{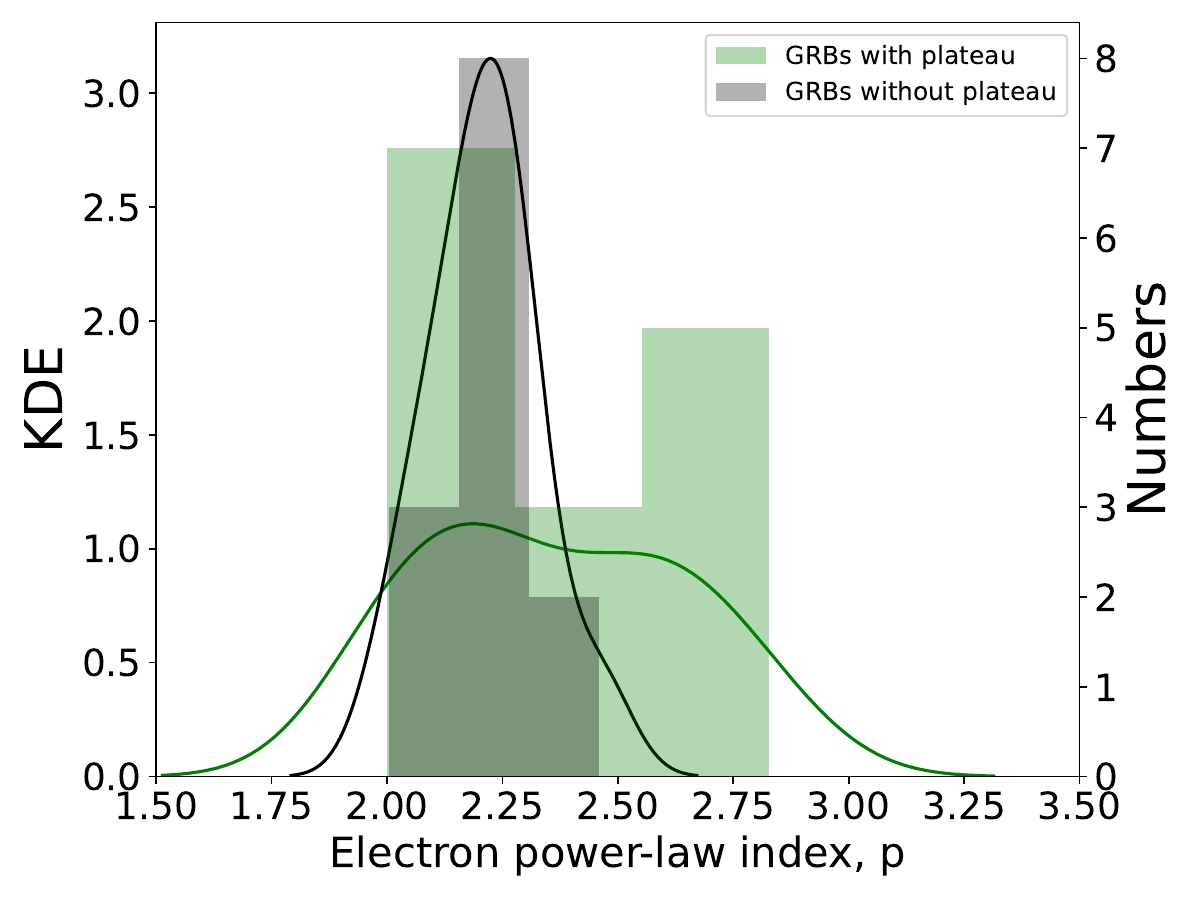}
    \caption{Distributions of the electron power-law index, $p$ (see Appendix A in \citealt{DB+25}) for GRBs with flares (left) and without flares (right). The color coding follows that of Figure~\ref{fig:time_after_early_steep_decay}. The corresponding number of GRBs in each subsample is listed in Table~\ref{tab:afterglow_statistics-1}. These distributions show no significant difference in the electron power-law index $p$ across the subsamples.}
    \label{fig:electron_power-law_index}
\end{figure}

\begin{figure}[H]
    \centering
    \includegraphics[width=0.45\linewidth]{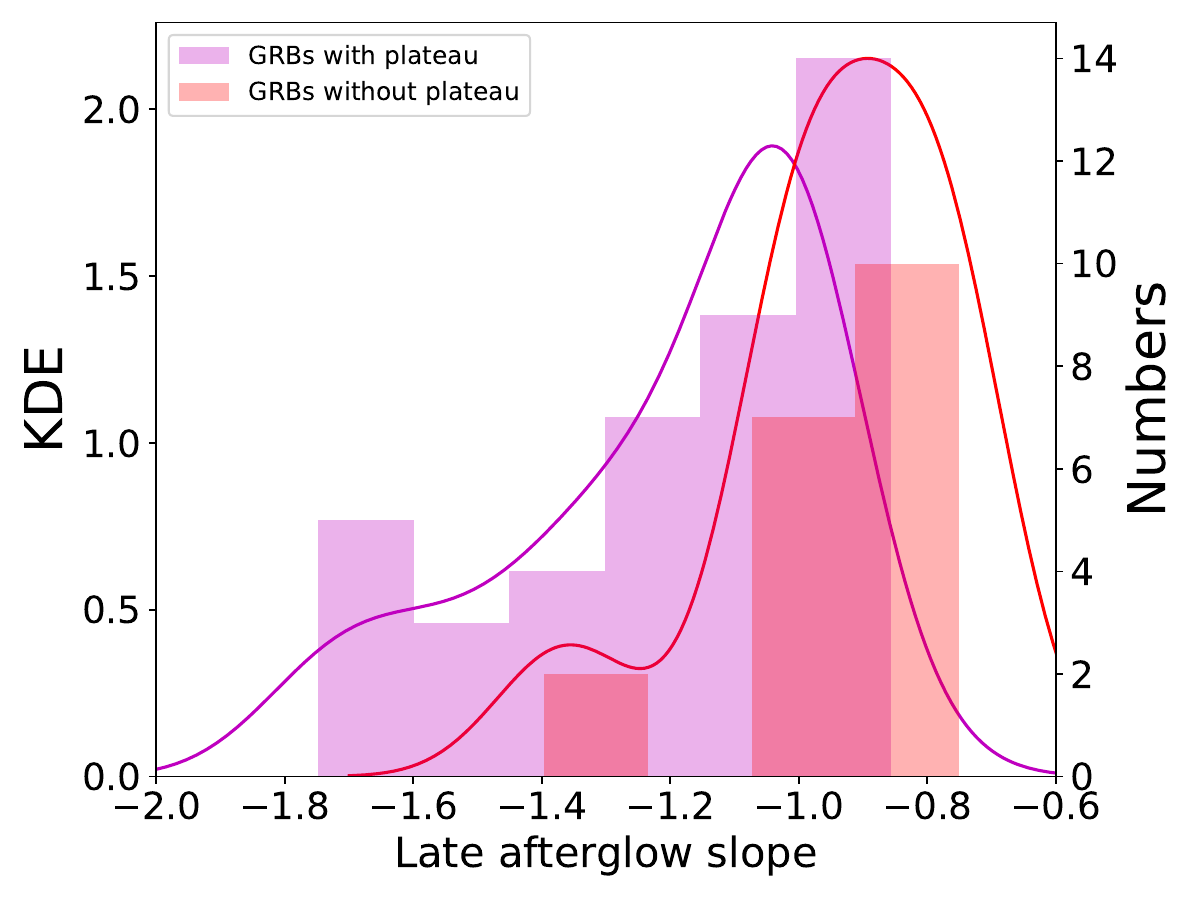}
    \includegraphics[width=0.45\linewidth]{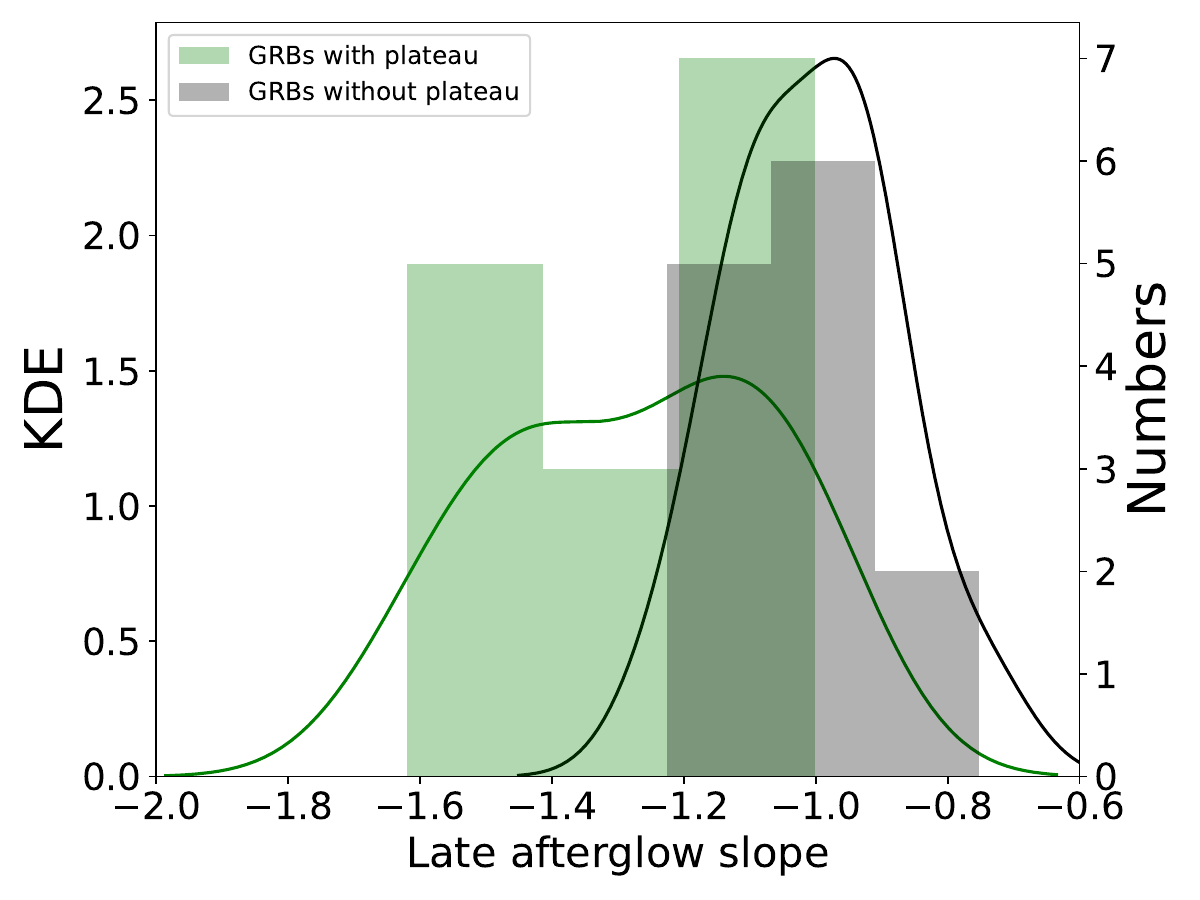}
    \caption{Distributions of the late afterglow slope obtained using electron power-law index, $p$ in different medium (see Appendix A in \citealt{DB+25}) for GRBs with flares (left) and without flares (right). The color coding follows that of Figure~\ref{fig:time_after_early_steep_decay}. The corresponding number of GRBs in each subsample is listed in Table~\ref{tab:afterglow_statistics-1}. These distributions show no significant difference in the late-time slope between the subsamples with and without flares. However, a clear distinction is observed between GRBs with and without a plateau, as confirmed by the K–S test result reported in Table~\ref{tab:afterglow_statistics-2} having $p < 0.05$.}
    \label{fig:late_afterglow_slope}
\end{figure}

\begin{figure}[H]
    \centering
    \includegraphics[width=0.45\linewidth]{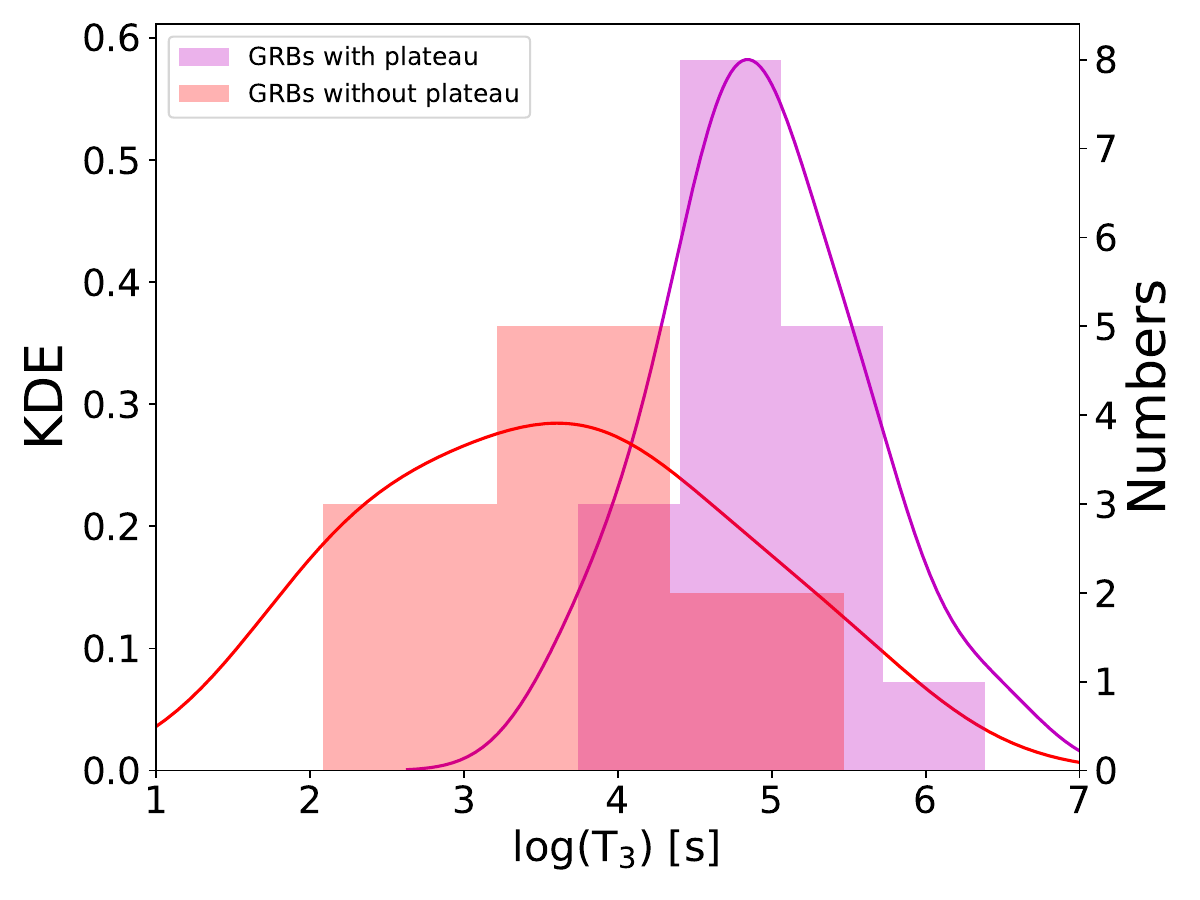}
    \includegraphics[width=0.45\linewidth]{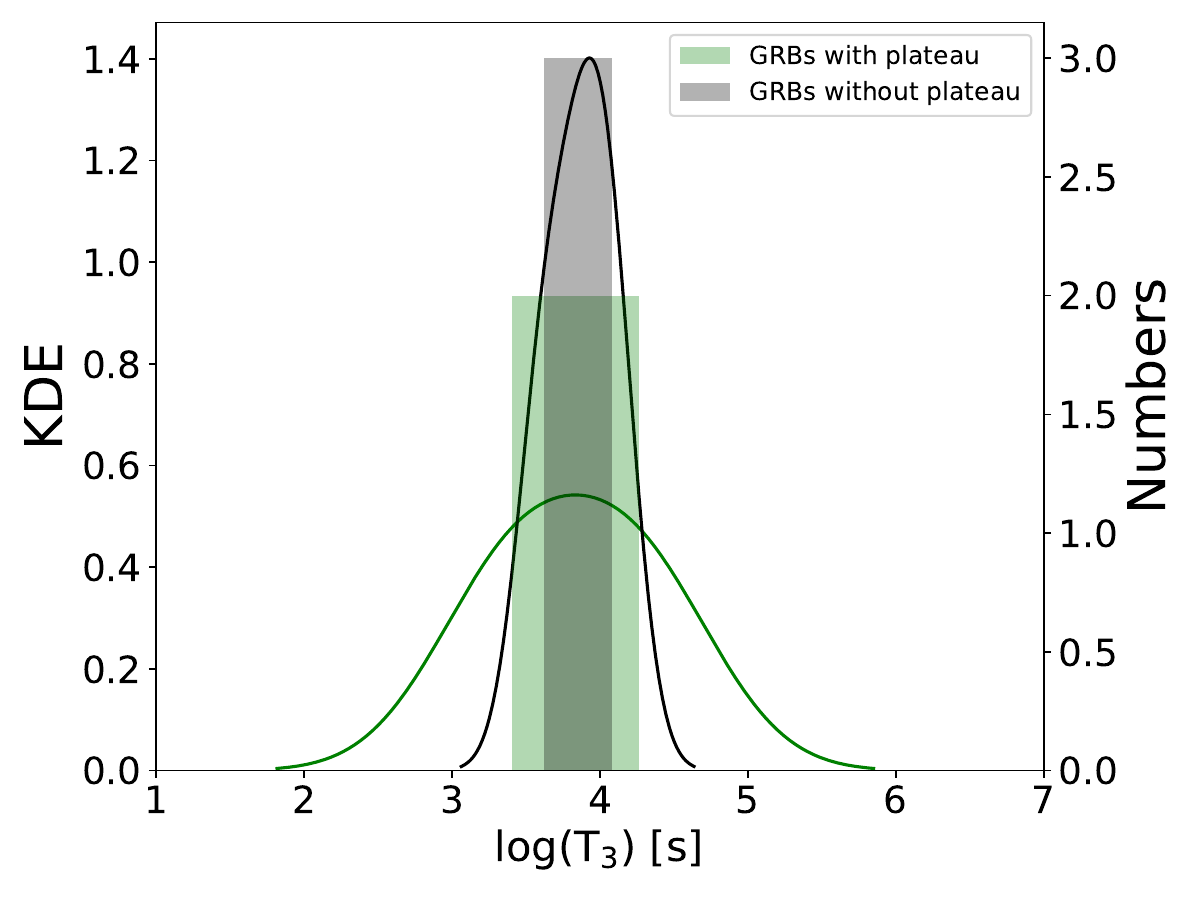}
    \caption{Distributions of the time of the jet break, $T_3$ (see Appendix A in \citet{DB+25}) for GRBs with flares (left) and without flares (right), with each panel showing the subsamples with and without plateaus. The color coding follows that of Figure~\ref{fig:time_after_early_steep_decay}. The corresponding number of GRBs in each subsample is listed in Table~\ref{tab:afterglow_statistics-1}.
    Since the number of GRBs without flares that show a jet break is limited, it is difficult to draw firm conclusions about statistical similarities or differences in $T_3$ across the subsamples. A statistically significant difference (1.01 days vs. 1.11 hours) is found only between GRBs with and without a plateau when flares are present.} 
    \label{fig:jet_break_time}
\end{figure}

The afterglow properties of GRB subsamples with and without flares, compared to those with and without plateaus, are presented in four figures. (i) The time at the end of the steep decay, which marks the transition to the afterglow phase (regular decay or ``plateau", if exists), denoted by  $T_1$ (Figure~\ref{fig:time_after_early_steep_decay}); (ii) the distribution of the electron power-law index $p$ \footnote{In our analysis, p is a fit parameter as the decay slopes of some of the afterglow segment can be expressed as a function of p, see \citet{DB+25}.} (Figure~\ref{fig:electron_power-law_index}), as fitted from the temporal evolution of the X-ray light curve after the steep decay phase;  (iii) the late-time (after the end of the plateau, if exists) afterglow slope (Figure~\ref{fig:late_afterglow_slope}) deduced using the electron power-law index $p$; and (iv) the jet break time $T_3$ (Figure~\ref{fig:jet_break_time}). These underlying afterglow parameters are directly obtained from the fit, as explained in \citet[][see Section 2.3]{DB+25}, and are listed in Appendix Table~\ref{tab:fit_params1}.
The statistical results for these parameters, including the number of GRBs, average values, and Kolmogorov–Smirnov (K-S) test outcomes, are summarized in Tables~\ref{tab:afterglow_statistics-1} and ~\ref{tab:afterglow_statistics-2}.

The distributions of the end time of the steep decay, $T_1$, presented in Figure~\ref{fig:time_after_early_steep_decay}, show consistent values across all four subsample types, with an average value of $\sim 321$ seconds. If one considers that the prompt emission lasts until this time due to high-latitude emission or curvature effect, this provides a clear indication that the mechanism causing this break is not affected by the presence of either a plateau or flares. Moreover, for the same subsamples we obtained the distributions of the electron power-law index $p$  from the observed light curve using the best fitted model; for details see 
 \citet{SPN98, DPR22}. The results are
presented in Figure \ref{fig:electron_power-law_index}, and show consistent values of $\sim 2.25$ (Table~\ref{tab:afterglow_statistics-1}). This result supports the interpretation that the power-law acceleration mechanism of the outflow is similar in both wind and ISM environments.

The distributions of the late-time afterglow slope, derived from the electron power-law index $p$, are presented in Figure~\ref{fig:late_afterglow_slope}. The results indicate distinct behaviors between subsamples with and without plateaus. The average values ($\sim -1.20$ with a plateau and $\sim -0.97$ without a plateau obtained from Table~\ref{tab:afterglow_statistics-1}) differ significantly, and the K–S test result of $p < 0.05$ reported in Table~\ref{tab:afterglow_statistics-2} confirms that they originate from different populations. As discussed in \citet{DPR22}, such a difference is expected when comparing wind and ISM environments. By contrast, the presence of flares has no effect on these distributions.

The distributions of the jet break time are presented in Figure~\ref{fig:jet_break_time}.
The results show a very pronounced distinction (by an order of magnitude) between jet break times for GRBs with and without a plateau when flares are present; in the absence of flares, the statistics are too limited to draw firm conclusions (a total of 5 GRBs, 2 with and 3 without plateaus). For GRBs with flares, the different average values ($\langle \log_{10} T_3 \rangle = 4.94 \pm 0.15$ and $3.60 \pm 0.34$ for bursts with and without a plateau, respectively, see Table~\ref{tab:afterglow_statistics-1}), together with the K–S test yielding $p < 0.05$ (see Table~\ref{tab:afterglow_statistics-2}), support this result. 
This distinction might be explained by different environments for GRBs with/without plateaus, e.g., ``wind" vs. ``ISM". We will present a detailed study of this result in future work. 
 
\begin{table}[H]
\centering
\caption{Statistics of the afterglow properties 1: Number of GRBs and average values with standard error of the mean (SEM) for afterglow properties in different subsamples. The parameters we show are 
(i) the end time of the steep decay, $T_1$, (ii) the time at the end of the plateau phase $T_{\rm a}\equiv T_2$, (ii) the jet break time $T_3$, (iv) the electron power-law index $p$, and (v) the late-time afterglow slope respectively. Average values indicate that while  $T_1$, $T_2$ and $p$ show no significant differences, the late-time afterglow slope and jet break time differ significantly between GRBs with and without a plateau. The presence of flares has no significant effect on the underlying afterglow parameters. For the jet break time ($T_{3}$), however, it is difficult to draw firm conclusions due to the small number of GRBs (2 with and 3 without plateaus) without flares.}
\begin{tabular}{l|cc|cc|cc|cc}
\toprule
 & \multicolumn{4}{c|}{With Flares} & \multicolumn{4}{c}{Without Flares} \\
\cmidrule(lr){2-5} \cmidrule(lr){6-9} 
& \multicolumn{2}{c|}{With Plateau} 
& \multicolumn{2}{c|}{Without Plateau} 
& \multicolumn{2}{c|}{With Plateau} 
& \multicolumn{2}{c}{Without Plateau} \\
\cmidrule(lr){2-3} \cmidrule(lr){4-5} \cmidrule(lr){6-7} \cmidrule(lr){8-9}
\textbf{Parameter}
 & $N_{\rm GRB}$ & Avg $\pm$ SEM
 & $N_{\rm GRB}$ & Avg $\pm$ SEM
 & $N_{\rm GRB}$ & Avg $\pm$ SEM
 & $N_{\rm GRB}$ & Avg $\pm$ SEM \\
\midrule
$ \log_{10} (T_{1}) $  & 42 & 2.59$\pm$0.07 & 7 & 2.59$\pm$0.18 & 15  & 2.46$\pm$0.13 & 7 & 2.34$\pm$0.18  \\
$ \log_{10} (T_2 \equiv T_{\rm a} ) $    & 42 & 4.03$\pm$0.14 & ... & ... & 15 & 3.73$\pm$0.21 & ... & ... \\
$ \log_{10} (T_{3}) $  & 17 & 4.94$\pm$0.15 & 10 & 3.60$\pm$0.34 & 2 & 3.86$\pm$0.30 & 3 & 3.83$\pm$0.30 \\
$\rm p$                & 42 & 2.27$\pm$0.05 & 19 & 2.15$\pm$0.05 & 15 & 2.21$\pm$0.03 & 13 & 2.36$\pm$0.07 \\
$\rm Afterglow~slope$  & 42 & -1.20$\pm$0.04 & 19 & -0.94$\pm$0.04 & 15 & -1.27$\pm$0.05 & 13 & -1.00$\pm$0.05 \\
\bottomrule
\end{tabular}
\label{tab:afterglow_statistics-1}
\vspace{5mm}
\end{table}

\begin{table}[H]
\centering
\caption{Statistics of the afterglow properties 2: Kolmogorov--Smirnov (K--S) test results, reporting the K--S statistic $D$ and the associated probability $p$ for afterglow properties in different subsamples. The parameters are described in Table \ref{tab:afterglow_statistics-1}. The K-S test results indicate that while $p$, $T_1$, and $T_2$ show no significant differences, the late-time afterglow slope and jet break time differ significantly ($p < 0.05$) between GRBs with and without a plateau. The presence of flares has no significant effect on the underlying afterglow parameters.}
\begin{tabular}{l|cc|cc|cc|cc}
\toprule
 & \multicolumn{2}{c|}{With Flares} & \multicolumn{2}{c|}{Without Flares} & \multicolumn{2}{c|}{With Plateau} & \multicolumn{2}{c}{Without Plateau} \\
\cmidrule(lr){2-3} \cmidrule(lr){4-5} \cmidrule(lr){6-7} \cmidrule(lr){8-9}
& \multicolumn{2}{c|}{With/Without Plateau} 
& \multicolumn{2}{c}{With/Without Plateau}
& \multicolumn{2}{c|}{With/Without Flares} 
& \multicolumn{2}{c}{With/Without Flares} \\
\cmidrule(lr){2-3} \cmidrule(lr){4-5} \cmidrule(lr){6-7} \cmidrule(lr){8-9}
\textbf{Parameter} & D & p & D & p & D & p & D & p \\
\midrule
$ \log_{10} (T_{1}) $   & 0.38 & 0.30 & 0.31 & 0.62 & 0.24 & 0.49& 0.43 & 0.58 \\
$ \log_{10} (T_2 \equiv T_{\rm a} ) $     & ...  & ...  & 0.20 & 0.72  & 0.20 & 0.72 & ... & ... \\
$ \log_{10} (T_{3}) $   & 0.68 & 0.003 & 0.50 & 0.90  & 0.88 & 0.07 & 0.50 & 0.49 \\
$\rm p$                 & 0.27 & 0.24 & 0.40 & 0.16 & 0.30 & 0.22& 0.50 & 0.03 \\
$\rm Afterglow~slope$   & 0.64 & 0.00001 & 0.66 & 0.002  & 0.30 & 0.22 & 0.40 & 0.14 \\
\bottomrule
\end{tabular}
\label{tab:afterglow_statistics-2}
\end{table}

\subsubsection{Properties of plateau with and without flares}

\begin{figure}[H]
    \centering
    \includegraphics[width=0.45\linewidth]{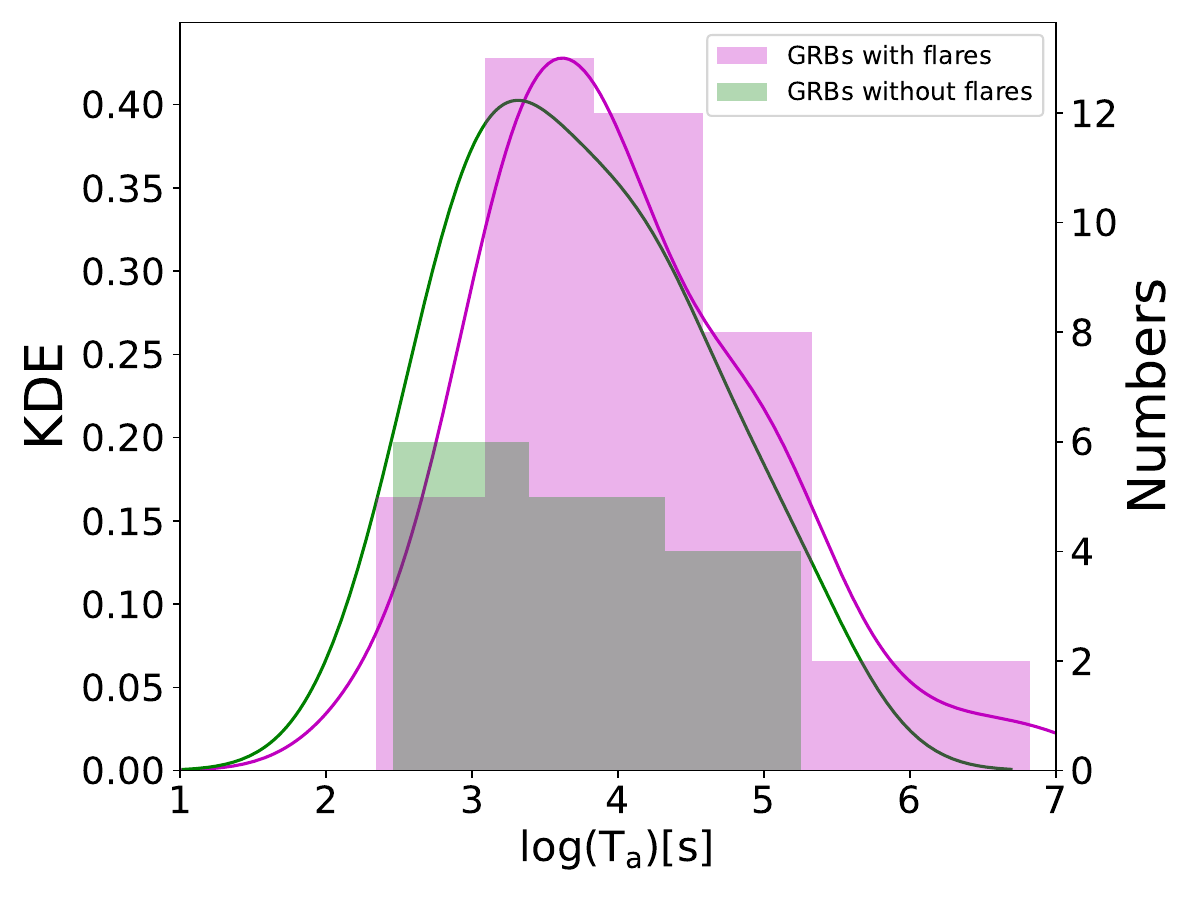}
    \caption{Distributions of the time at the end of the plateau phase $T_{a}$ for GRBs with (purple) and without (green) flares. The 42 GRBs with plateau phase and flares are in purple, while the 15 GRBs with plateaus phase but without flares are in green. These distributions show no significant difference at $T_a \equiv T_2$ between the GRBs with and without flares.}
    \label{fig:plateu_end_time}
\end{figure}

Figure \ref{fig:plateu_end_time} shows the distribution of time at the end of the plateau phase, $T_{\rm a} \equiv T_2$ (see Appendix A in \citet{DB+25} for detailed discussion).
According to our interpretation \citep{DPR22}, this time marks the end of the coasting phase of the GRB outflow, and the transition of the blast-wave to the self-similar phase.  
The average of the distribution is $\left < \log_{10} T_{\rm a}\right > = 3.95 \pm 0.12$ for all 57 GRBs with a plateau phase. When separating the sample to those with and without flares, as shown in Figure \ref{fig:plateu_end_time}, we find no clear distinction between the two subsamples. Comparing the averages, we find that $\left < \log_{10} T_{\rm a}\right > = 4.03 \pm 0.14$ for 42 GRBs with flares and $\left < \log_{10} T_{\rm a}\right > = 3.73 \pm 0.21$ for 15 bursts without flares (these results are also presented in Table~\ref{tab:afterglow_statistics-1}). We further performed  a K-S test for both subsamples and found that the test statistic is $D = 0.20$ with a probability $p = 0.72$ for $T_{\rm a}$ (these results are also presented in Table~\ref{tab:afterglow_statistics-2}). Since $p > 0.05$, we cannot reject the null hypothesis at the 5\% significant level, suggesting that the two subsamples originate from the same population. We conclude that the occurrence of the flares has no effect on the end time of the plateau phase.

To summarize the main results of this section, we find that regardless of the presence or absence of flares, the distributions of all investigated afterglow parameters appear similar across subsamples as shown in Figures~\ref{fig:time_after_early_steep_decay}–\ref{fig:plateu_end_time}. This is also reflected in the average values summarized in Table~\ref{tab:afterglow_statistics-1}. In particular, the electron power-law index $p$ shows consistent values ($\sim 2.25$) regardless of flare presence, supporting the interpretation that flares are indeed additional emission components superimposed on the underlying afterglow continuum, rather than compensations for model limitations. Moreover, the afterglow break times $T_1$ (the end time of the steep decay) and $T_{\rm a} \equiv T_2$ (the time at the end of the plateau phase)
are found to be similar across all sample types, with average values of 321 seconds and $10^4$ seconds
respectively. 
Furthermore, the results of the K-S tests (see Table \ref{tab:afterglow_statistics-2}) indicate that the distributions of each parameter in the subsamples are statistically consistent with originating from the same population. These distributions demonstrate that flare properties are independent of the presence of a plateau phase, and thus of the underlying afterglow properties in GRB X-ray light curves. These findings are consistent with the results presented in Section~\ref{sec:flare_properties} and in \citet{DB+25}, where no significant differences were found in the distributions of flare parameters between GRBs with and without plateaus.


\section{Prompt analysis results: GRBs with/without flares and plateaus}
\label{sec:prompt_properties}

The prompt properties of the GRB subsamples with and without flares, compared to those with and without plateaus, are presented in Figures~\ref{fig:hist_t90_Eiso_Epeak_flares_plateaus}–\ref{fig:hist_redshift_flares_plateaus} and summarized in Appendix Table~\ref{tab:prompt_parameters1}. Figure~\ref{fig:hist_t90_Eiso_Epeak_flares_plateaus} shows the distributions of burst duration $T_{90}$, isotropic energy $E_{\rm iso}$, and peak energy $E_{\rm pk}$. The computations of $E_{\rm iso}$ and $E_{\rm pk}$ are described in Appendices~\ref{app:Eiso_computation} and \ref{app:Epk_computation}, respectively. Figure~\ref{fig:hist_spectral_index_flares_plateaus} presents the distribution of the spectral index, obtained primarily from Swift-BAT and, when available, from Fermi-GBM, derived from the best-fit spectral model. Finally, Figure~\ref{fig:hist_redshift_flares_plateaus} displays the redshift distributions for the different subsamples.

The primary statistical analysis indicates that the prompt properties of the GRB subsamples are generally consistent regardless of the presence of a plateau phase. This trend is also reflected in the average values summarized in Table~\ref{tab:prompt_statistics-1}. Furthermore, the results of the K–S tests presented in Table~\ref{tab:prompt_statistics-2} (in rows 2-3 and 4-5 respectively, when testing with/without plateaus presence) indicate that the distributions of each parameter in the subsamples are statistically consistent with originating from the same population. Nevertheless, small differences exist, which we highlight and present below. 

For the peak energy $E_{\rm pk}$ (see Figure~\ref{fig:hist_t90_Eiso_Epeak_flares_plateaus}, top right), the distributions of GRBs without a plateau phase tend to peak at higher values, and the trend becomes even clearer when considering the subsample without flares (see Figure~\ref{fig:hist_t90_Eiso_Epeak_flares_plateaus}, bottom right). To highlight this tendency, we plot a dashed black vertical line at around 300 keV in each $E_{\rm pk}$ distribution.  This is similar to a difference we observed in our previous study between GRBs with and without a plateau phase (see Fig. 6 in \citealt{DPR22}). In that study, the GRBs with a plateau phase included some events classified as low-luminosity GRBs, based on both their intrinsic properties \citep{GDV07} and their X-ray afterglow emission \citep{Dereli+17}, while the GRBs without a plateau phase were mostly, if not entirely, observed by Fermi-LAT, which tends to detect the brightest GRBs. However, a population-based study is needed to determine whether such a difference truly exists between low- and high-luminosity GRBs.

In addition, the spectral index distributions shown in Figure~\ref{fig:hist_spectral_index_flares_plateaus} display some differences across the subsamples. 
GRBs with a plateau phase tend to have a softer spectral index, a trend that becomes even more pronounced when considering the subsample without flares and supported with moderate K-S test ($D = 0.44$, $p = 0.1$) presented in Table \ref{tab:prompt_statistics-2}. This behavior resembles that seen in the $E_{\rm pk}$ distributions (Figure~\ref{fig:hist_t90_Eiso_Epeak_flares_plateaus} (right), discussed above), where the same subsamples with plateau also tends to peak at lower energies compare to the subsamples without a plateau \footnote{Part of this softening may reflect the fitting method: single power-law fits tend to yield softer photon indices than the low-energy slopes of Band or cutoff power-law models, since they approximate the spectral curvature near $E_{\rm pk}$, especially when $E_{\rm pk}$ lies within or below the instrument’s band.}. The soft spectral indices are consistent with nonthermal origin of the prompt phase such as synchrotron emission \citep[e.g.,][]{Oganesya+2018, YDR19, Burgess+2020, DPR20}. We therefore conclude that GRBs with a plateau phase show a tendency toward softer spectra and lower $E_{\rm pk}$ values, compared to the harder spectra and higher energies of those without a plateau. This independent result further suggests that thermal emission is unlikely to be present in GRBs with plateau phases, consistent with the predictions of our model presented in \citet{DPR22}. Therefore, the presence or absence of a plateau phase does not significantly affect the prompt properties of GRBs with or without flares.

When the same test is applied to the presence or absence of flares, we find that $\left < \log_{10} T_{90} \right >$ tends to be higher for GRBs with flares compared to those without, as shown in Tables~\ref{tab:prompt_statistics-1} and \ref{tab:prompt_statistics-2}. A similar trend is also observed for $\left < \log_{10} E_{\rm iso} \right >$ and $\left < \log_{10} E_{\rm pk} \right >$. This result is also confirmed when comparing the average values of each prompt parameter distribution presented in Table~\ref{tab:prompt_statistics-1} with those of all studied 89 GRBs in Appendix~\ref{app:prompt_properties}. This might indicate that GRBs with flares are generally more energetic, peak at higher energies, and last longer than those without flares. Such a conclusion is supported by the findings and discussion in Section~\ref{sec:abundance_of_X-ray_flares}, where the existence of a jet break time during the afterglow was found to be more common in GRBs with flares, suggesting that these bursts may indeed be more energetic and longer-lasting. 
Moreover, as noted in the Swift-BAT GRB catalog\footnote{\url{https://swift.gsfc.nasa.gov/results/batgrbcat/}}, almost all GRBs without flares and without a plateau exhibit short-duration, single-peaked light curves, in contrast to GRBs with flares, which typically display multiple peaks and last longer.

\begin{figure}
\centering
\includegraphics[width=0.32\linewidth]{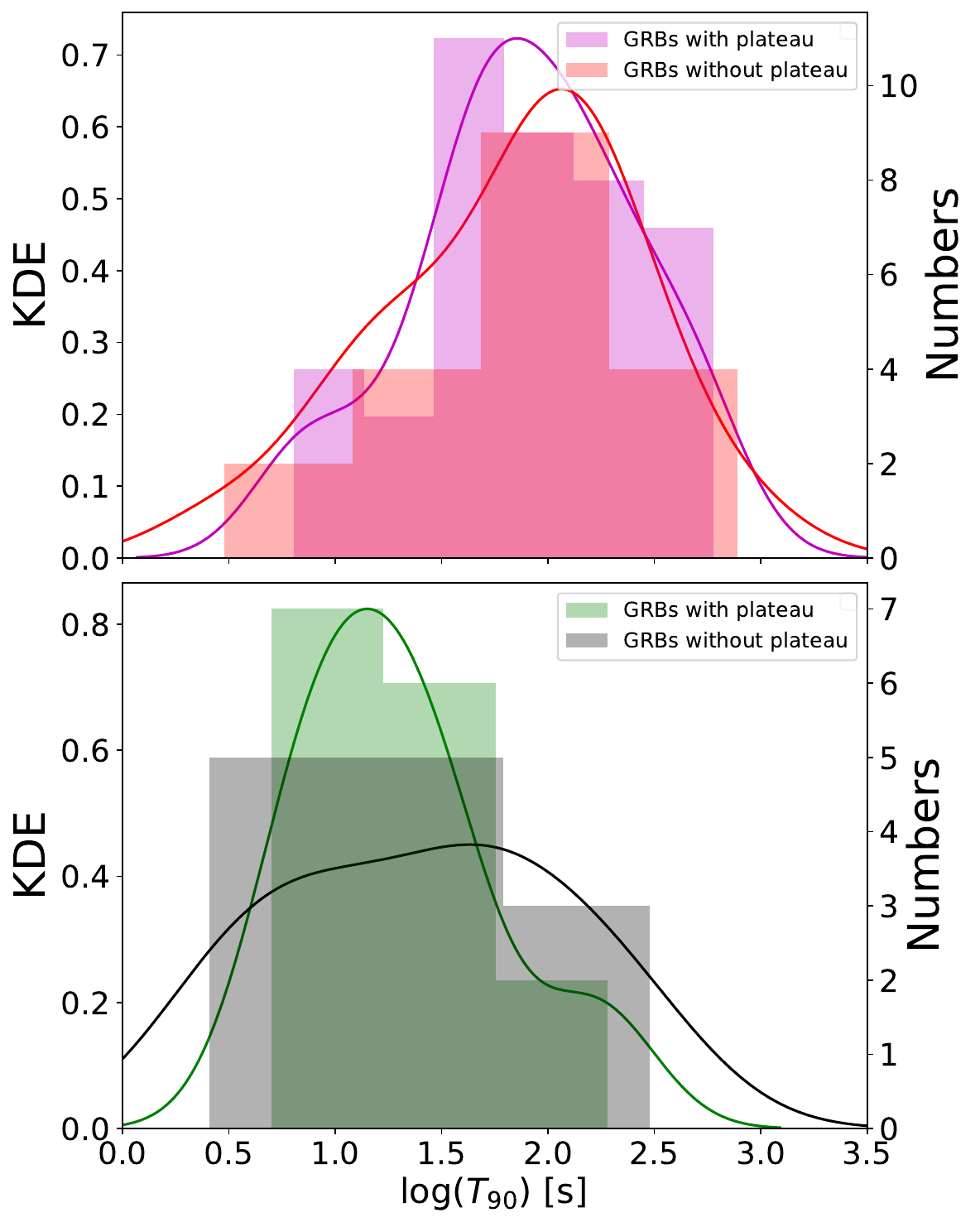}
\includegraphics[width=0.32\linewidth]{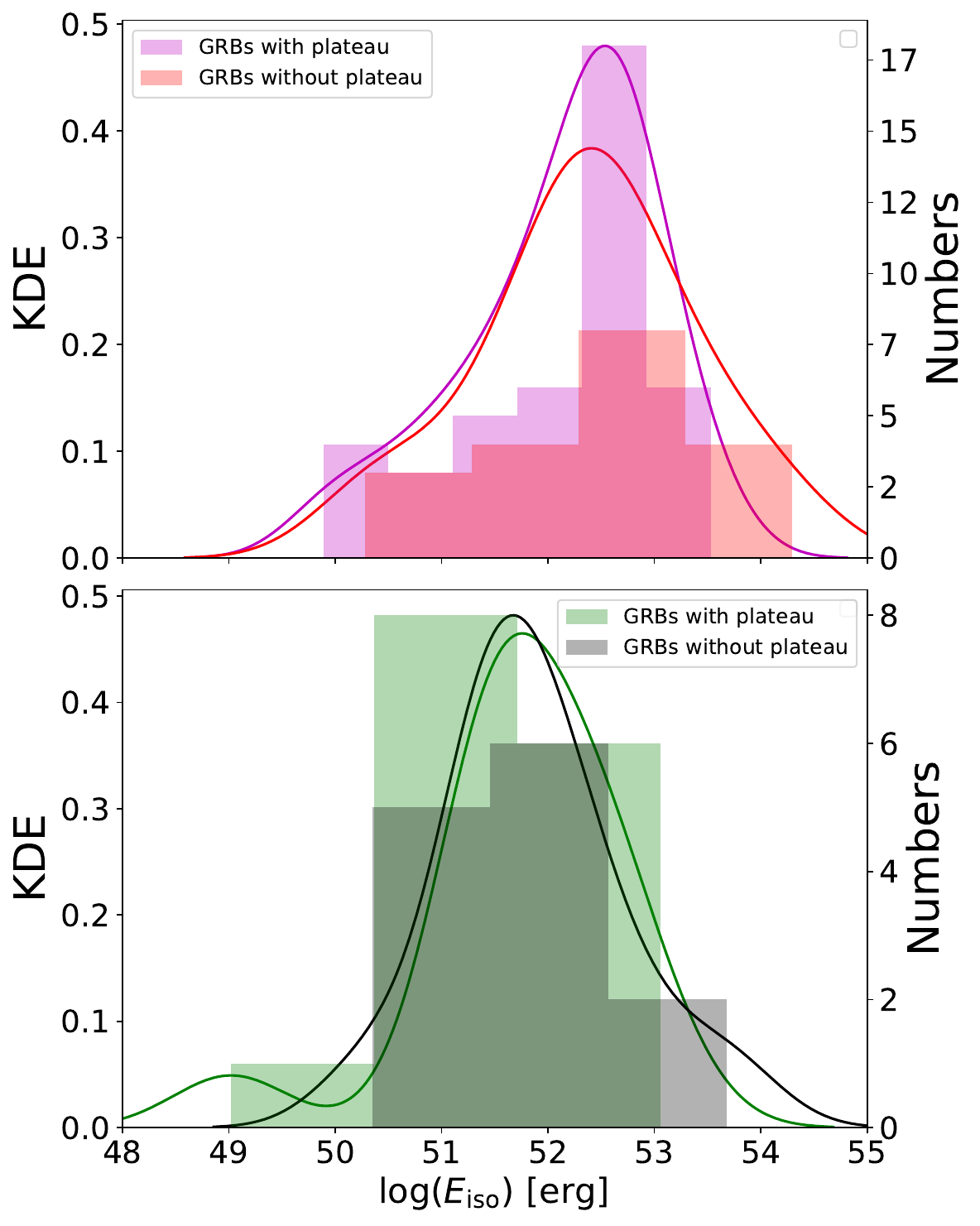} 
\includegraphics[width=0.32\linewidth]{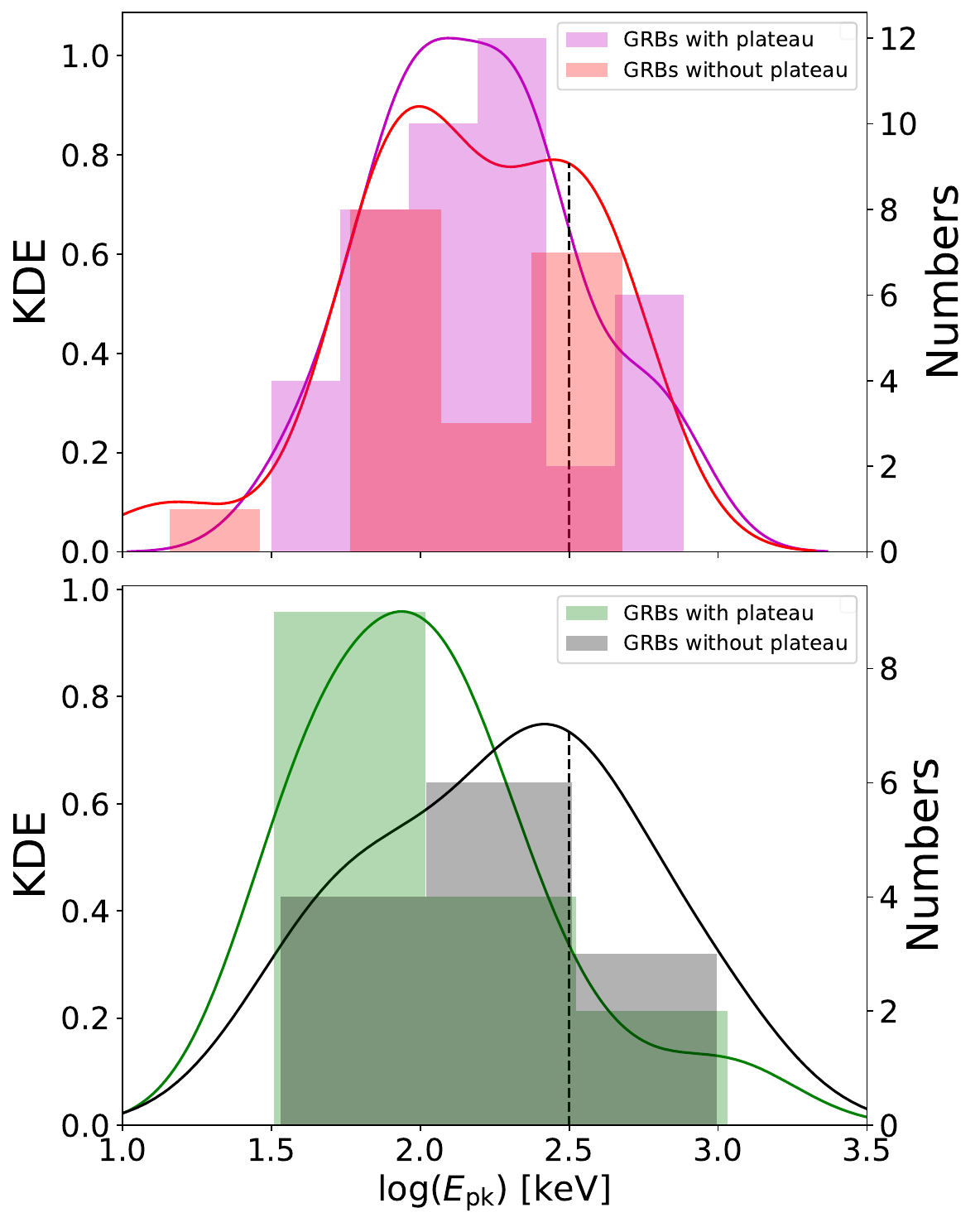}
\caption{Distributions of prompt properties: burst duration $T_{90}$ (top and bottom left), isotropic energy $E_{\rm iso}$ (top and bottom middle), and peak energy $E_{\rm pk}$ (top and bottom right), shown for GRBs with and without flares, with each panel showing the subsamples with and without plateaus. The color coding follows that of Figure~\ref{fig:time_after_early_steep_decay}. The corresponding number of GRBs in each subsample is listed in Table~\ref{tab:prompt_statistics-1}, in the relevant column. These distributions (except for the peak energy $E_{\rm pk}$) show no significant difference across the subsamples. 
The $E_{\rm pk}$ distributions of GRBs without a plateau phase and without flares tend to peak at higher values (bottom right), with a marginal similar tendency for the $E_{\rm pk}$ distribution of GRBs with flares (top right); the dashed vertical black line added at 300 keV indicates this tendency. The results show that the prompt properties of the GRBs with and without flares are the same, regardless of whether there is a plateau phase in the GRBs X-ray light curves. 
}
\label{fig:hist_t90_Eiso_Epeak_flares_plateaus}
\end{figure}

To conclude, we find that the prompt emission properties, $T_{90}$, $E_{\rm pk}$, $E_{\rm iso}$, as well as spectral slopes and light curve shapes, are generally similar for GRBs with and without plateau phases, except small spectral variations seen in the corresponding distributions and discussed above. However, notable differences are observed when comparing GRBs with and without flares. In particular, GRBs with flares tend to exhibit longer durations, and greater isotropic energies.
They also display more complex, multi-peaked light curves. These differences may hint at the existence of distinct GRB populations, possibly associated with low- and high-luminosity regimes. Therefore, while the presence of a plateau phase does not appear to significantly affect the prompt properties, the presence of flares correlates with more energetic and longer-lasting prompt emission. 

\begin{figure}
\centering
\includegraphics[width=0.45\linewidth]{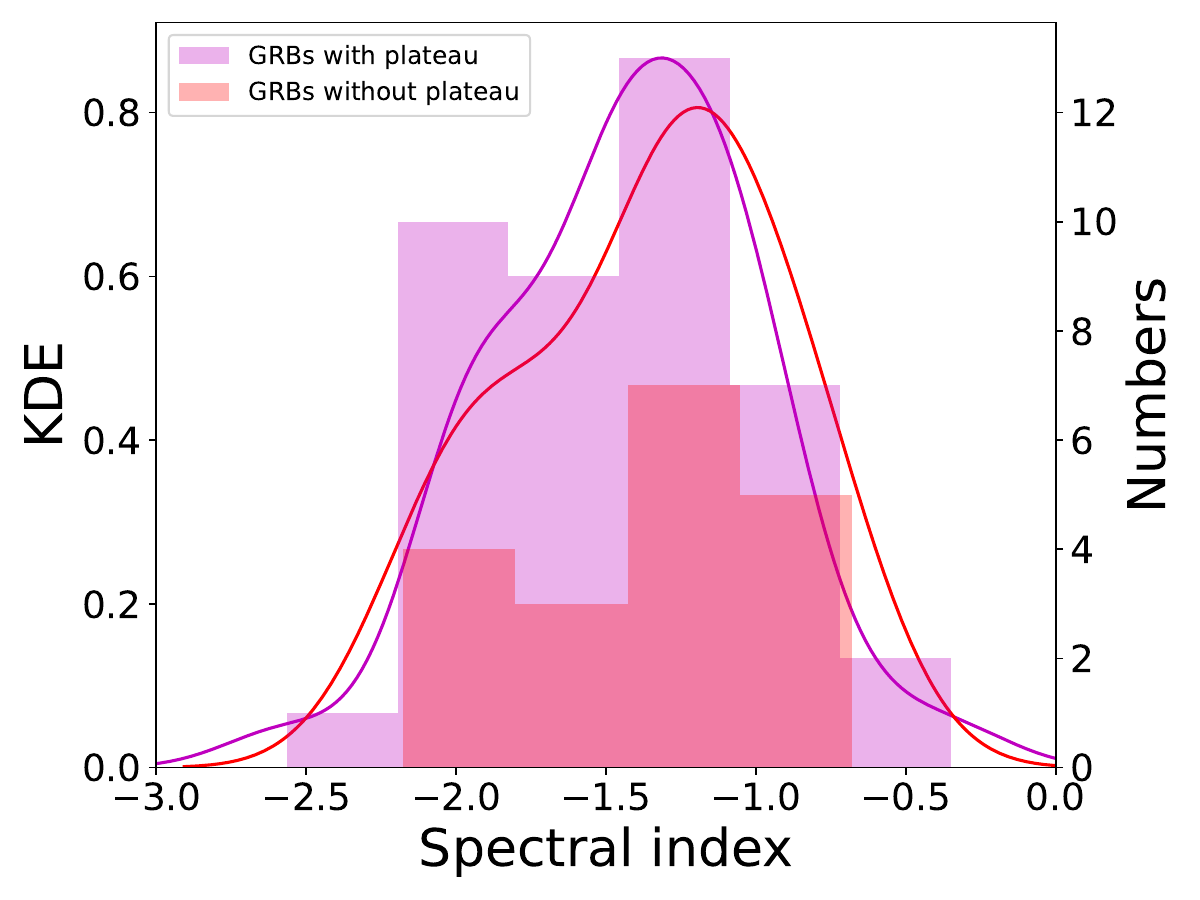}
\includegraphics[width=0.45\linewidth]{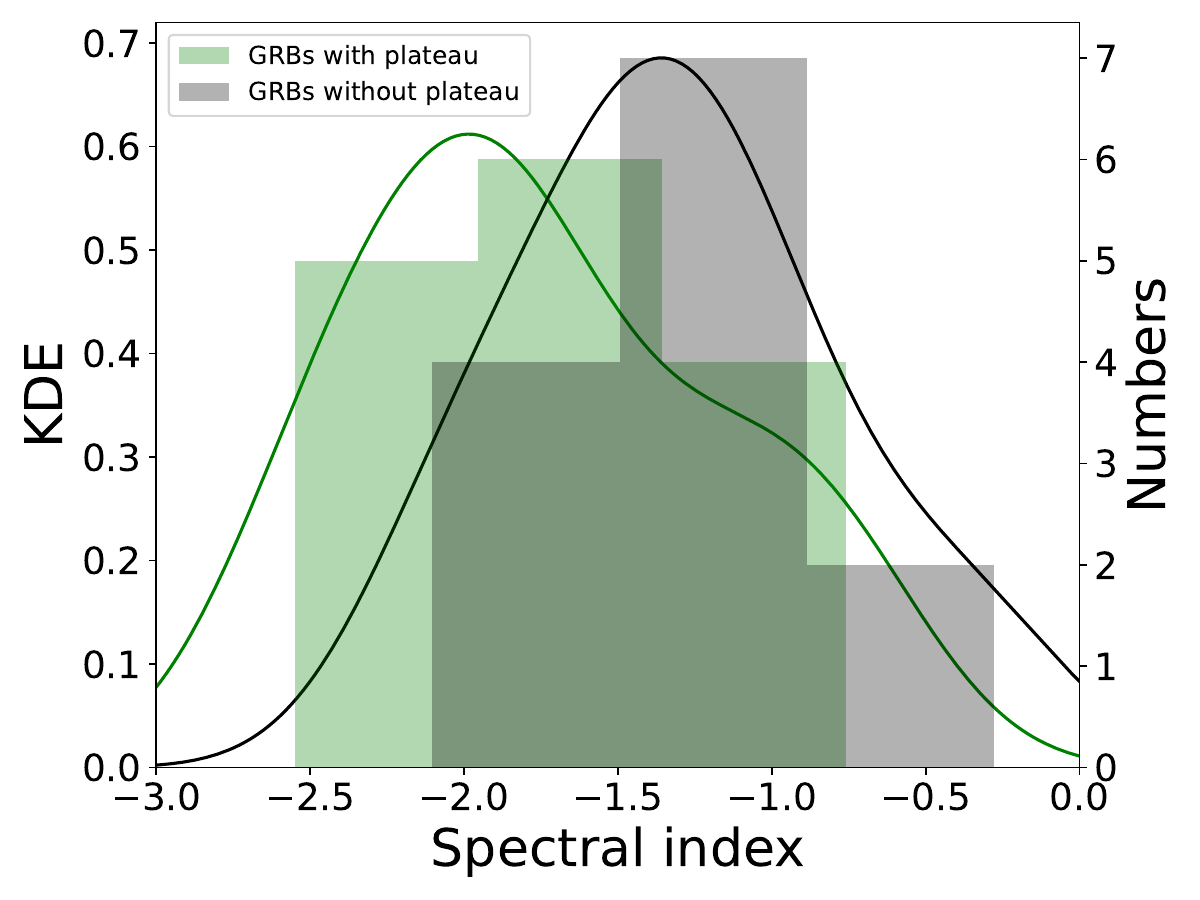}
\caption{Distributions of the spectral index obtained from different best fitted model parameters from two instruments (Swift-BAT and Fermi-GBM) as explained in Appendix \ref{app:Eiso_computation}: for GRBs with flares (left) and without flares (right), with each panel showing the subsamples with and without plateaus. The color coding follows that of Figure~\ref{fig:time_after_early_steep_decay}. The corresponding number of GRBs in each subsample is listed in Table~\ref{tab:prompt_statistics-1}, in the relevant column. These distributions show a trend toward a softer spectral index for GRBs with a plateau phase, which is even more pronounced for GRBs without flares.} 
\label{fig:hist_spectral_index_flares_plateaus}
\end{figure}


\begin{figure}
\centering
\includegraphics[width=0.45\linewidth]{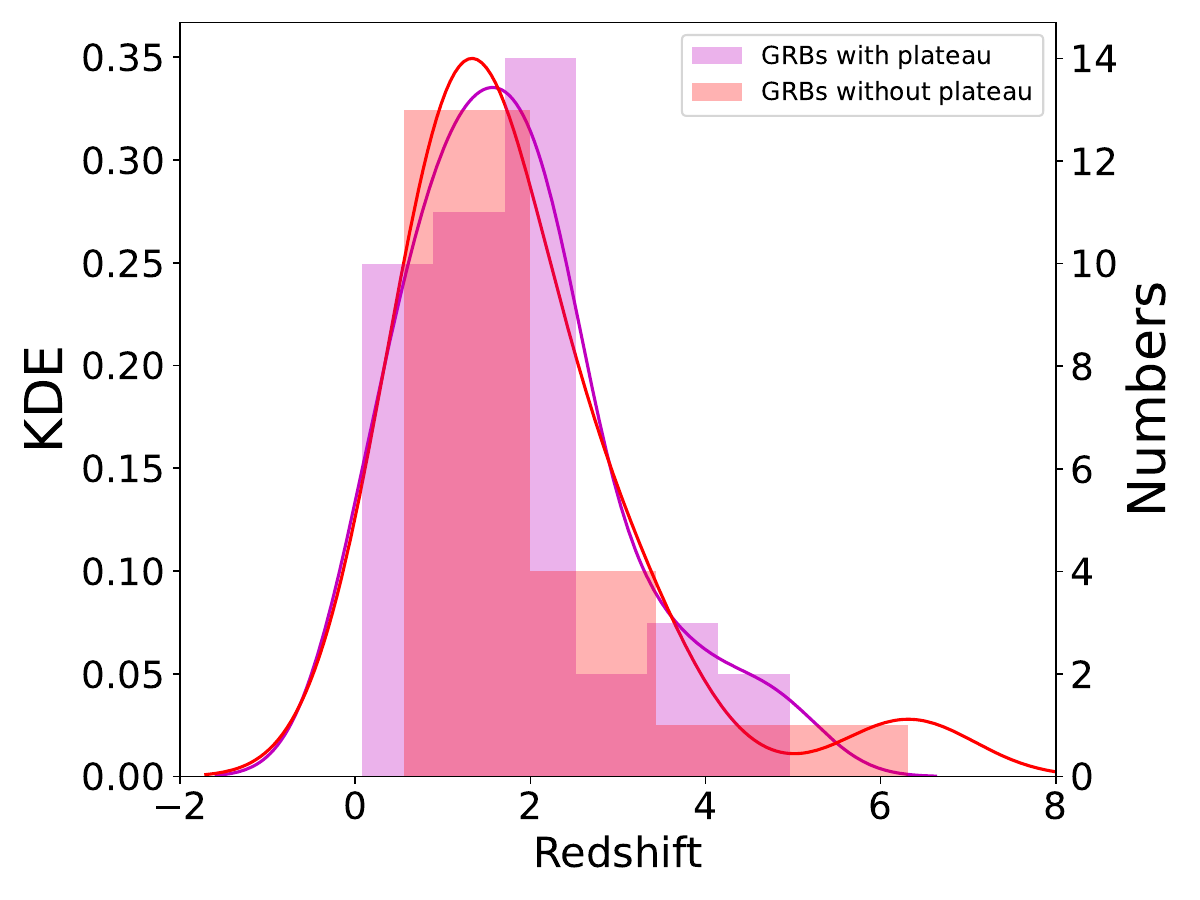}
\includegraphics[width=0.45\linewidth]{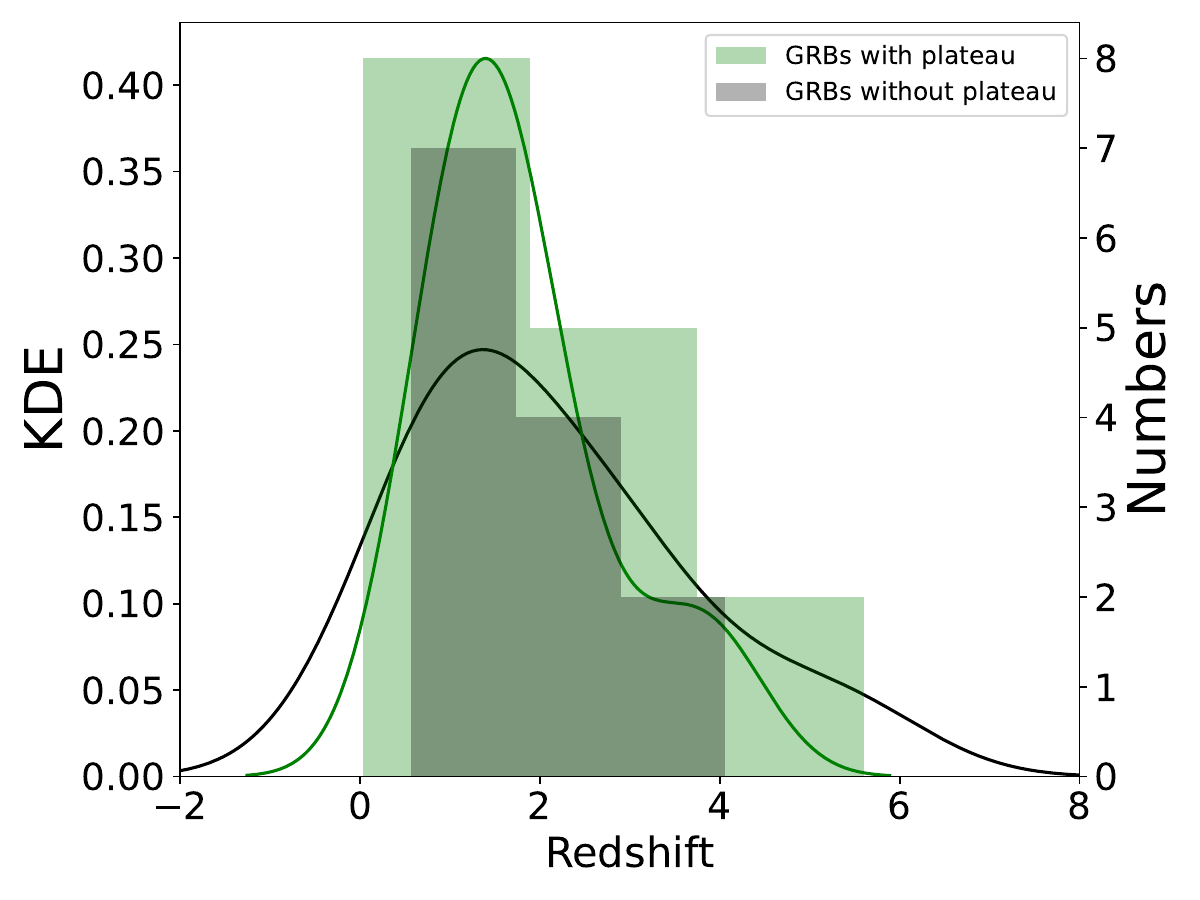}
\caption{Distributions of the redshift for GRBs with flares (left) and without flares (right), with each panel showing the subsamples with and without plateaus. The color coding follows that of Figure~\ref{fig:time_after_early_steep_decay}. The corresponding number of GRBs in each subsample is listed in Table~\ref{tab:prompt_statistics-1}, in the relevant column. These distributions show no significant differences across the subsamples.} 
\label{fig:hist_redshift_flares_plateaus}
\end{figure}

\begin{table}[H]
\centering
\caption{Statistics of the prompt properties 1: Number of GRBs and average values with standard error of the mean (SEM) for prompt properties in different subsamples. The parameters shown are burst duration $T_{90}$, isotropic energy $E_{\rm iso}$, peak energy $E_{\rm pk}$, spectral index, and redshift $z$. Average values indicate no significant differences between GRBs with and without plateaus.}
\begin{tabular}{l|cc|cc|cc|cc}
\toprule
 & \multicolumn{4}{c|}{With Flares} & \multicolumn{4}{c}{Without Flares} \\
\cmidrule(lr){2-5} \cmidrule(lr){6-9}
& \multicolumn{2}{c|}{With Plateau } 
& \multicolumn{2}{c|}{Without Plateau } 
& \multicolumn{2}{c|}{With Plateau } 
& \multicolumn{2}{c}{Without Plateau } \\
& \multicolumn{2}{c|}{(42 GRBs)} 
& \multicolumn{2}{c|}{ (19 GRBs)} 
& \multicolumn{2}{c|}{ (15 GRBs)} 
& \multicolumn{2}{c}{ (13 GRBs)} \\
\cmidrule(lr){2-3} \cmidrule(lr){4-5} \cmidrule(lr){6-7} \cmidrule(lr){8-9}
\textbf{Parameter} 
 & & Avg $\pm$ SEM
 & & Avg $\pm$ SEM
 & & Avg $\pm$ SEM
  & Avg $\pm$ SEM \\
\midrule
$ \log_{10} (T_{90}) $   &  & 1.89$\pm$0.08 & & 1.83$\pm$0.13 & & 1.30$\pm$0.12  & 1.41$\pm$0.18 \\
$ \log_{10} (E_{iso}) $  &  & 52.12$\pm$0.14 & & 52.36$\pm$0.23  & & 51.74$\pm$0.23  & 51.86$\pm$0.22 \\
$ \log_{10} (E_{pk}) $  &   & 2.19$\pm$0.05 & & 2.16$\pm$0.09  & & 2.01$\pm$0.10  & 2.28$\pm$0.12 \\
Spectral index         &   & -1.43$\pm$0.07 & & -1.37$\pm$0.10 & & -1.31$\pm$0.14  & -1.75$\pm$0.14 \\
Redshift ($z$)        &    & 1.78$\pm$0.18 & & 1.86$\pm$0.30  & & 2.12$\pm$0.39  & 1.80$\pm$0.27 \\
\bottomrule
\end{tabular}
\label{tab:prompt_statistics-1}
\end{table}

\begin{table}[H]
\centering
\caption{Statistics of the prompt properties 2: Kolmogorov--Smirnov (K--S) test results, reporting the K--S statistic $D$ and the associated probability $p$ for prompt properties in different subsamples. The parameters are described in Table \ref{tab:prompt_statistics-1}.  K-S test results show no significant differences between GRBs with and without a plateau. However, the results when plateau present suggest that GRBs with flares tend to have longer durations, higher energies and softer spectra compared to those without flares. While we cannot claim that the distributions are drawn from different populations due to $p > 0.05$, the moderate K–S statistics for $T_{90}$, $E_{iso}$, $E_{pk}$ and spectral index support these trends. However, there is no difference in the redshift distributions between the two subsamples. In the subsample without plateaus, none of the parameters showed statistically significant differences between GRBs with and without flares.}
\begin{tabular}{l|cc|cc|cc|cc}
\toprule
 & \multicolumn{2}{c|}{With Flares} & \multicolumn{2}{c|}{Without Flares} & \multicolumn{2}{c|}{With Plateau} & \multicolumn{2}{c}{Without Plateau} \\
\cmidrule(lr){2-3} \cmidrule(lr){4-5} \cmidrule(lr){6-7} \cmidrule(lr){8-9}
& \multicolumn{2}{c|}{With/Without Plateau} 
& \multicolumn{2}{c}{With/Without Plateau}
& \multicolumn{2}{c|}{With/Without Flares} 
& \multicolumn{2}{c}{With/Without Flares} \\
\cmidrule(lr){2-3} \cmidrule(lr){4-5} \cmidrule(lr){6-7} \cmidrule(lr){8-9}
\textbf{Parameter} & D & p & D & p & D & p & D & p \\
\midrule
$ \log_{10} (T_{90}) $   & 0.15 & 0.88 & 0.33 & 0.35 & 0.68 & 0.00002 & 0.40 & 0.13 \\
$ \log_{10} (E_{iso}) $  & 0.16 & 0.81 & 0.19 & 0.92 & 0.34 & 0.13 & 0.43 & 0.10 \\
$ \log_{10} (E_{pk}) $   & 0.18 & 0.73 & 0.40 & 0.13 & 0.32 & 0.17 & 0.22 & 0.77 \\
Spectral index           & 0.20 & 0.57 & 0.44 & 0.10 & 0.41 & 0.04 & 0.17 & 0.94 \\
Redshift ($z$)           & 0.16 & 0.83 & 0.25 & 0.70 & 0.23 & 0.51 & 0.13 & 0.99 \\
\bottomrule
\end{tabular}
\label{tab:prompt_statistics-2}
\vspace{5mm}
\parbox{\textwidth}{\small }
\end{table}


\section{Distinguishing X-ray flares from afterglow and prompt properties}
\label{sec:discussion}

\subsection{Implication of X-ray plateau presence for the X-ray flare properties}

The results presented here support and add to the findings of our previous work \citep{DB+25}, where we showed that the presence or absence of a plateau phase does not influence the observed characteristics and properties of flares, such as rise and decay times  (see Section \ref{sec:correlations2}).
This result provides key constraints on the theoretical models proposed to explain the plateau phase, a feature seen in approximately 60\% of GRB X-ray light curves. 

This shallow decay phase, which deviates from standard afterglow predictions, has long posed a challenge for theoretical models. Among the leading scenarios invoked are: (i) late-time energy injection into the blast wave \citep{Nousek06, Metzger11, DallOsso+2011, Cannizzo+2011, vanEerten2014a, vanEerten2014b, Ronchini+2023, Lenart+2025}; (ii) the viewing-angle effects, where the observer’s line of sight lies outside the main jet cone \citep{EG06, Eichler14, BDD+20a}; and (iii) the coasting phase of a jet with a relatively low Lorentz factor \citep{ShM12, DPR22}. The first two scenarios predict differences in flare timing and structure that are not supported by the data. In particular, late energy injection should result in delayed or narrower flares in some cases, while an observer viewing the jet off-axis should lead to systematically delayed flare times due to reduced Doppler boosting, neither of which are observed and deduced from our analysis. By contrast, the low Lorentz factor model aligns with the observations: flares originating from smaller radii (as expected in lower-$\Gamma$ jets) would still be observed at similar times due to time-dilation effects, yielding comparable observed flare properties regardless of plateau presence. These findings offer independent support for the hypothesis that the plateau phase may appear only in GRBs with relatively low Lorentz factors, on the order of a few tens as previously proposed in \citet{DPR22}. 

\subsection{Abundance of X-ray flares across GRB subsamples}
\label{sec:abundance_of_X-ray_flares}
We show that GRBs with flares are as abundant in plateau bursts ($73\%$) as in the general GRB population. In subsection \ref{sec:statistical_results}, we report that flares appear in approximately $69\%$ of all GRBs (61 GRBs), indicating that flares are quite common in GRB afterglows. This value is slightly higher than previously reported, where about half of GRBs were found to exhibit flares \citep{Zhang06, Chincarini+10, Racusin+16}. We demonstrate that the difference in the flare fraction compared to previous studies arises primarily from the time window considered for flare detection. Unlike earlier works, we include all flares, from those observed at the tail end of the prompt emission to both early- and late-time flares. When excluding GRBs with flares occurring at the tail end of the prompt emission, we are left with  $63\%$ of all GRBs that have flares characterized by $T_{90} < t_{\rm pk}$. When further excluding late-times, one finds that $60\%$ of all GRBs have flares with $t_{\rm pk} < 1000$ s. 
When applying both conditions simultaneously, namely $T_{90} < t_{\rm pk} < 1000$~s, we find that the flare fraction decreases to $52\%$, consistent with previous results. 
Furthermore, a recent study by \citet{Ma&Gao+2025}, using a fitting approach similar to \citet{Chincarini+10}, showed that separating flares occurring during the prompt phase ($t_{\rm pk}<T_{90}$) from later flares does not significantly affect the distributions of their properties. This supports our conclusion that flare behavior is consistent across all temporal regimes, independent of their timing. 

Applying the same cut ($T_{90} < t_{\rm pk} < 1000$ s) to identify how many GRBs with flares also exhibit plateaus, we find that flares remain nearly as common in plateau bursts as in the full GRB sample: 54\% vs. 52\%, respectively. This confirms that the presence of flares is independent of the presence of plateaus, even when excluding flares occurring at the tail end of the prompt emission and beyond 1000 s.

As presented in section~\ref{sec:statistical_results}, 80\% of all 89 GRBs in our sample exhibit a steep decay, while a jet break is observed in 36\%. The ratio of steep decays remains consistent in GRBs with flares (80\%) and in those without flares (79\%), which means that the existence of flares has no effect on the steep decay. 
However, jet breaks are more frequently identified in GRBs with flares (44\%) compared to those without (18\%), though this may partly reflect observational biases, as flare-producing GRBs tend to be brighter and more extensively monitored, increasing the likelihood of detecting late-time features such as jet breaks.

When separating the sample by plateau presence, we find that steep decays are ubiquitous in plateau GRBs, occurring in 100\% of cases both with and without flares, while their frequency is lower in non-plateau bursts (37\% with flares, 54\% without). This pattern is tightly linked to the canonical light curve morphology and appears largely independent of the presence of flares. 
In contrast, jet break rates are less common overall and relatively consistent across all subsamples: 41\% (with plateau) and 53\% (without plateau) in GRBs with flares, and 13\% vs. 23\% in those without flares. We thus conclude that, as opposed to the steep decays, jet breaks are largely unaffected by existence of flares or plateaus, and are therefore controlled by an external parameter, namely the jet opening angle.

\subsection{Implications of X-ray flares and plateau presence for the underlying afterglow properties}

Our analysis of the afterglow parameters presented in Section \ref{sec:afterglow_properties} show no significant differences between GRBs with and without flares. This consistency strongly suggests that the presence of flares does not depend on variations in the underlying afterglow properties. This reinforces the idea that the mechanism responsible for flares is likely distinct from that producing the underlying afterglow emission.

Comparisons between X-ray flare properties and late-time afterglow characteristics have been explored in a few previous studies. For instance, \citet{Curran+08}, using a small sample of eight GRBs, found no correlation between the timing and flux variability of flares and the underlying temporal decay slope (see their Fig.~2). Similarly, \citet{Racusin+16} showed that the presence of flares or plateau phases does not affect the positive correlation between the X-ray luminosity at 200 seconds and the afterglow decay index thereafter. However, the more comprehensive comparison presented in this work, examining both the properties and presence/absence of flares in relation to the full set of afterglow parameters, has not been carried out in previous studies. Our findings thus provide the most statistically robust evidence to date that the occurrence of flares is decoupled from the global dynamics of afterglow evolution. 

Furthermore, we show that the existence of a plateau phase does not affect any of the early underlying afterglow properties. However, the presence of a plateau influences the late-time behavior. We find that the late-time afterglow slopes and jet break times differ significantly between GRBs with and without plateaus: the slope differences are independent of the presence of flares, whereas the jet break times diverge only when flares are present, due to the small number of GRBs available without flares (2 with and 3 without a plateau). This result is consistent with the broader picture in which GRBs with plateaus are associated with wind environments, whereas those without plateaus are linked to ISM environments. In this interpretation, the later jet break times naturally follow from the dependence of the break time on the density profile of the external medium. These findings provide independent confirmation that our model selection and fitting procedure are unbiased and correctly applied to the data. A more detailed exploration of the role of the environment in shaping jet break times will be presented in future work.

\subsection{Implications of X-ray flares and plateau presence for the prompt emission properties}
\label{prompt_emission_Insights}

We investigated in Section \ref{sec:prompt_properties} whether the presence of a plateau phase or X-ray flares in GRB afterglows is associated with differences in the prompt emission properties. We found that they remain nearly unchanged between GRBs with and without a plateau phase. This suggests that the occurrence of a plateau in the X-ray afterglow does not correlate with significant differences in the energetics or temporal structure of the prompt phase.  However, some mild trends are observed. In particular, GRBs without a plateau phase tend to show slightly higher $E_{\rm pk}$ values and harder spectra compared to those with plateaus. This tendency becomes more pronounced in GRBs without both a plateau and flares. A possible interpretation, as discussed in \citet{DPR22}, is that the GRBs without a plateau phase may preferentially belong to a higher-luminosity population which are mostly detected by Fermi-LAT, whereas GRBs with plateaus include some events associated with low-luminosity populations \citep{GDV07, Dereli+17}. 
A comparison of the spectral index of the prompt emission reported in \citet{SIM23} with our overlapping sample (5 GRBs with plateaus versus 1 GRB without) suggests the same spectral variation as in our analysis: GRBs without plateaus tend to have a harder spectral index compared to those with. However, this sample is small, therefore, a dedicated population study would be needed to confirm whether this difference is intrinsic or observational/instrumental effect. 
The increased low-energy sensitivity of the recently launched SVOM mission, together with CTAO’s forthcoming capability to probe the high-energy afterglow, and thereby help distinguish high-luminosity from low-luminosity GRB populations, will enable tests of whether the harder prompt spectra of non-plateau GRBs correspond to a high-luminosity population and whether the softer spectra of plateau GRBs indicate a low-luminosity one.

When examining the role of flares, the data suggests that GRBs with flares tend to have slightly longer durations and higher isotropic energies compared to those without flares. This is reflected in the average values ($\langle \log T_{90} \rangle = 1.89 \pm 0.08$ vs. $1.30 \pm 0.12$ with a plateau and $1.83 \pm 0.13$ vs. $1.41 \pm 0.18$ without a plateau, and $\langle \log E_{\rm iso} \rangle = 52.12 \pm 0.14$ vs. $51.74 \pm 0.23$ with a plateau and $52.36 \pm 0.23$ vs. $51.86 \pm 0.22$ without a plateau, all is cgs units; see Table~\ref{tab:prompt_statistics-1}) and supported by moderate K–S statistics ($D = 0.68, 0.40$ with a probability $p = 2\times10^{-5}, 0.13$ for $T_{90}$ and $D = 0.34, 0.43$ with a probability $p=0.13, 0.10$ for $E_{\rm iso}$ with and without a plateau respectively, see Table~\ref{tab:prompt_statistics-2}). In addition, GRBs with flares tend to display more complex, multi-peaked light curves. This indicates that GRBs with flares may be, on average, more energetic and longer-lasting. 

Nonetheless, we note that the differences seen in the prompt parameter distributions in the presence of flares are not that significant as can be seen by the K-S statistics in Table \ref{tab:prompt_statistics-2}, except $T_{90}$ for bursts with plateau. And in addition, they may be influenced by selection effects or small-number statistics and systematic effects arising from different instruments (Swift-BAT versus Fermi-GBM), as we also include Fermi-GBM data when available. Therefore, the results should be interpreted with caution, especially since the distributions do not include errors on the computed parameters of the prompt properties. Indeed, a recent independent study by \citet{SIM23} showed that the properties of the prompt emission obtained from Swift-BAT data are not affected by the presence of flares when considering parameter distributions alone. Therefore, this remains an open question, particularly when considering all the possible biases. 

In terms of redshift distribution, we find no notable difference between the subsamples. This supports the conclusion that the observed variations in prompt properties are not due to redshift-driven selection biases. 

We conclude that the presence of flares may be associated with more energetic and longer-lasting prompt emission, whereas the existence of a plateau phase appears to affect only the spectral properties. Specifically, we find a tendency toward softer spectra and lower $E_{\rm pk}$ values for GRBs with a plateau phase, compared to the harder spectra and higher energies of those without. However, this spectral variation requires further dedicated study, particularly in light of the possible biases discussed above.


\section{Implication of the flare asymmetry and parameter correlations on the origin of the flares} 
\label{sec:asymmetry_implication}
Our analysis in Section \ref{sec:correlations2} shows that X-ray flares are highly asymmetric in both GRB subsamples, with and without a plateau phase, with decay times approximately five times longer than rise times. The average $t_{\rm rise}/t_{\rm decay}$ ratios of 0.2 and 0.19 we find are notably smaller than previously reported values for prompt pulses (\( \langle t_{\rm rise}/t_{\rm decay} \rangle = 0.47 \)) and early X-ray flares (\( \langle t_{\rm rise}/t_{\rm decay} \rangle = 0.49 \pm 0.26 \)) \citep[e.g.,][]{KRL03, Chincarini+10}, suggesting a more pronounced asymmetry in our analysis. This pronounced asymmetry likely arises from two factors: (i) our temporal fits are performed in log-space rather than in linear space, as in earlier studies, and (ii) we isolate the flare component from the underlying afterglow using physically motivated models. The pronounced asymmetry therefore reflects an intrinsic property of the flare source. 

In addition, we find a strong positive correlation between the rise and decay times, consistent across both GRB subsamples ($r = 1.3 \pm 1.6 \times 10^{-6}$ for bursts with a plateau and $r = 0.66 \pm 1.2 \times 10^{-6}$ for those without). 
This correlation suggests a consistent scaling behavior between the flare rise and decay phases, even if the underlying physical processes may differ. 
Such a relationship can be interpreted as evidence of self-similar behavior in the flare profiles \citep{Bernardini+11}.

In our analysis, we find  $ \langle t_{\rm rise}/t_{\rm pk} \rangle \simeq 1/4$. Furthermore, in our previous work we showed that the average ratio of flare width to peak time is $\langle w/t_{\rm pk} \rangle \sim 1.7$ \citep{DB+25}. While these results are aligned with the predictions from the curvature effect of a relativistic expanding shell \citep{Kumar&Panaitescu2000, IKZ05, Liang+06, LP07}, the fact that the flares behave independent of the afterglow, points to a different origin (see below).

The asymmetry, relative rise time or width, and the correlation between rise and decay times appear to be independent of the presence of a plateau phase. 
This indicates that the underlying flare mechanism operates in a similar manner across different GRB afterglow environments. 
To our knowledge, this finding has not been highlighted before and supports the view that the temporal structure of flares is intrinsic to the emission process itself, rather than being shaped by external factors such as the viewing angle effects or external shocks. 

In addition, as presented in \citet{DB+25}, we observe a strong positive correlation between flare width and peak time across all subsamples, with Spearman’s coefficients of $r = 0.68$ and $r = 0.77$ (for GRBs with and without plateaus, respectively). Such a correlation was not seen in GRB prompt pulse analysis, reinforcing the view that X-ray flares represent a distinct emission component, physically separated from the prompt phase \citep{Ramirez-Ruiz+2000}; see, however recent works by \citet[][]{Hakkila+24, Gowri+25}, suggesting a possible continuity between late prompt pulses and flares. This result may originate from a refined definition of pulses \citep[see][for a detailed discussion]{Gowri+25}.

Taken together, these results point toward a self-similar and intrinsic nature of the flare emission, which is not significantly affected by external factors such as the circumburst environment or the viewing geometry. This strongly suggests that the origin of the flares lies in a long-lasting central engine activity.

Following the stellar collapse that triggers a GRB, a plausible scenario is that material from the progenitor star forms an accretion disk around the newly formed black hole. Matter from this disk gradually accretes onto the black hole, while instabilities within the disk lead to fluctuations in the accretion rate and the production of flares. The outflowing material associated with these flares exhibits self-similar dynamics that can persist much longer than the initial collapse episode producing the prompt GRB. 
Such a scenario is further supported by the clear distinction between the flare and afterglow behaviors, which effectively rules out an external origin for the flares.

Indeed, similar interpretations have been proposed in previous works, notably the accretion models of \citet{King+2005}, \citet{Perna+2006}, \citet{Proga&Zhang2006} and \citet{Kumar+2008a}. In this framework, late-time accretion instabilities naturally produce intermittent energy release, leading to flares with temporal characteristics consistent with those observed in our study. This supports the interpretation that flares may originate from late-time energy injection associated with accretion-disk instabilities at small radii and appear at late-times.

As we showed here, flares do not correlate with the presence of a plateau, and existence of a plateaus does not affect flare properties. These results therefore suggest that the origin of flares must be separated from the origin of the plateau. One can therefore conclude that late-time energy injection model cannot explain the origin of the plateau phase.


\section{Summary and Conclusion}
\label{sec:conclusion}

In this work, we compare the properties of X-ray flares with both prompt and afterglow properties by dividing the sample of 89 GRBs into four groups with and without flares and with and without plateaus.
Our analysis shows that approximately $69\%$ of all GRBs have flares. While $63\%$ of all GRBs have a plateau phase, $73\%$ of those with plateau also exhibit flares. From these results, we conclude that the existence of flares is independent of the existence of a plateau.

We find that about 80\% of GRBs in our sample exhibit steep decay phases and 36\% show jet breaks. Steep decays are independent of flares but strongly associated with plateaus, occurring in 100\% (with and without flares) of plateau GRBs versus only 37–54\% (with and without flares respectively) of non-plateau GRBs. In contrast, jet breaks are less frequent ($\sim$ 36\% overall) and occur at comparable rates across subsamples, reflecting their link to external shock dynamics rather than flare or plateau presence. 
These results emphasize that steep decays trace internal emission processes tied to the canonical light curve structure, while jet breaks unaffected by the presence of flares or a plateau phase, controlled by an external parameter, namely the jet opening angle.

When analyzing the properties of flares that were detected in those 61 GRBs (of those GRBs, 42 (68\%) have a plateau, while 19 do not), we found
no statistical difference in the distributions of the flare peak times, flare width, flare rise and decay times and the ratio of flare width to peak time, $w/t_{pk} \sim 1$ between GRBs with and without
plateau presented in Section \ref{sec:flare_properties}. From these results,  we conclude that the flare properties of GRBs are similar regardless of the presence
or absence of plateau phases in the GRBs X-ray light curves. This show that the flare properties are independent of afterglow characteristics, particularly those associated with the plateau phase. This result puts constraints on the models explaining the origin of the plateaus as discussed in \citet{DB+25}. 

Furthermore, by analyzing the rise and decay times of X-ray flares, we find that flares are highly asymmetric in both GRB subsamples, with and without a plateau phase. The average ratios of rise to decay time are  \( \langle t_{\rm rise}/t_{\rm decay} \rangle = 0.2 \) and \( \langle t_{\rm rise}/t_{\rm decay} \rangle = 0.19 \) respectively, indicating that decay times are typically five times longer than rise times. In addition to this pronounced asymmetry, we observe a strong positive correlation between rise and decay times across both subsamples ($r = 1.3$, $p = 1.6 \times 10^{-6}$ for GRBs with plateaus; $r = 0.66$, $p = 1.2 \times 10^{-6}$ for those without). These results appear independent of whether the flare is the first or second in the burst, and independent of the plateau phase, supporting the interpretation that the temporal structure of flares is intrinsic to the emission process itself, rather than shaped by external factors.
As discussed in Section \ref{sec:asymmetry_implication}, these results especially the pronounced asymmetry and self-similarity indicate that X-ray flares originate from intrinsic variability of the central engine, such as late-time accretion, rather than from external shock processes.

Our analysis reveals that the afterglow properties of GRBs with flares are statistically indistinguishable from those without flares. Specifically, parameters such as the electron power-law index ($p \sim 2.25$), break times ($T_1$ and $T_3$), and the late-time afterglow slope shown in Figures~\ref{fig:electron_power-law_index}–\ref{fig:jet_break_time} and summarized in Tables~\ref{tab:afterglow_statistics-1} and ~\ref{tab:afterglow_statistics-2}, exhibit no significant differences across subsamples. Additionally, the end time of the plateau phase ($T_{\rm a}$) (Figure~\ref{fig:plateu_end_time}) is consistent between GRBs with and without flares. These findings demonstrate that the presence of flares is not correlated with changes in the underlying afterglow dynamics. Consequently, we conclude that flares are likely produced by a separate mechanism, distinct from external forward shock emission and are instead superimposed on the standard afterglow light curve. To conclude, these distributions demonstrate that flare properties are independent of the presence of a plateau phase, and thus of the underlying afterglow properties in GRB X-ray light curves.

We find no statistically significant differences in prompt emission properties, such as burst duration ($T_{90}$) (except $T_{90}$ for bursts with plateau), isotropic energy ($E_{\rm iso}$), peak energy ($E_{\rm pk}$), or redshift, when dividing the sample based on the presence or absence of X-ray flares or plateau phases (Figures~\ref{fig:hist_t90_Eiso_Epeak_flares_plateaus}–\ref{fig:hist_redshift_flares_plateaus}; Tables~\ref{tab:prompt_statistics-1} and \ref{tab:prompt_statistics-2}). This supports the conclusion that neither the plateau phase nor the presence of flares in the X-ray afterglow is determined by global characteristics of the prompt phase. Nonetheless, mild trends are observed: GRBs with flares tend to be longer, more energetic, and often exhibit more complex, multi-peaked light curves, while GRBs without a plateau phase show a tendency toward harder spectra and higher $E_{\rm pk}$ values. These tendencies, although not statistically conclusive, may point to population-level differences, such as a bias toward high-luminosity bursts in the non-plateau sample, as previously suggested \citep{DPR22}. Finally, we find no difference in the redshift distributions across subsamples, ruling out redshift-driven selection effects. Overall, our results indicate that the presence of flares may be linked to more powerful prompt emission, whereas the existence of a plateau phase leaves little imprint on global prompt properties, though its possible role in shaping spectral variation warrants further study. 

Therefore, based on the results presented in Sections~\ref{sec:prompt_properties}, \ref{sec:flare_properties}, and \ref{sec:afterglow_properties}, we conclude that the prompt, flare, and afterglow properties of GRBs remain consistent regardless of the presence or absence of a plateau phase in their X-ray light curves. This places important constraints on the models proposed for both the origin of the plateau phase \citep{DB+25} and the flares. As discussed in Section \ref{sec:asymmetry_implication}, our results suggest that flares may originate from late-time energy injection powered by accretion-disk instabilities. As flare occurrence does not correlate with the presence of plateau, and vise versa, the presence of flares does not affect the underlying afterglow properties, particularly the existence of a plateau, the plateau must have a separate origin. Excluding the late time energy injection model, leaves the low Lorentz factor proposal \citep{DPR22} as a leading model for explaining the origin of the GRB X-ray plateau.

\begin{acknowledgments}
This work made use of data supplied by the UK Swift Science Data Center at the University of Leicester. AP, acknowledges support from the European Union via ERC consolidating grant $\sharp$773062 (acronym O.M.J.) and from the Israel Space Agency via grant number 6766. FR acknowledges support from the Swedish National Space Agency (2021-00180 and 2022-00205)
\end{acknowledgments}

\vspace{5mm}
\facility{Swift-XRT}
\software{MultiNest}

\bibliography{biblio_new}{}
\bibliographystyle{aasjournal}

\appendix
\restartappendixnumbering  

\section{Sample of 89 GRBs and afterglow best fit parameters} \label{app:GRBs_fit_params}
In our analysis, the fit parameters are computed using the maximum posterior estimate (MAP) method. The derived parameters and their errors are calculated using the marginalized posterior distributions (MPD). As discussed in \citet{DB+25} (Sec. 2.4) and summarized in Section \ref{sec:Fitting_procedure}, the selection of the best fit models is based on the AIC, and AICc criteria. The results show that the AICc condition is consistently satisfied by the data. Additionally, in most cases, the AIC condition is also met with the same model. Therefore, we only present the AICc in Table \ref{tab:fit_params1}, along with the some best fit parameters and some derived parameters (flare rise and decay times) obtained from the best fit parameters which are used in this study. We also want to refer the Table C1 in \citet{DB+25} for the rest of the derived parameters used in this study: flare peak time, flare width, flare asymmetry, total flare isotropic energy, flare width to the flare peak time ratio, and flare peak flux-to-underlying continuum ratio. The parameter errors are computed by using the credible intervals (e.g., 68\% for $1\sigma$) derived directly from the posterior samples of the fit parameters.

\begin{deluxetable}{cccccccccc}[H]
\tabletypesize{\scriptsize}
\tablecaption{Best fit parameters for 89 GRBs (Part 1). Column 1: GRB name, Column 2: Best model, Column 3: Corrected Akaike Information Criterion of the best model. Columns 4 and 5: Some derived flare parameters with their errors; flare rise and decay times. Note that when the best model has two flares, the entry of flare rise and decay times corresponding to this GRB is on two rows. Columns 6-10: Best fit parameters with their errors; (i) the end time of the steep decay $T_1$, (ii) the time at the end of the plateau phase $T_{\rm a}\equiv T_2$, (iii) the jet break time $T_3$ (iv) the electron power-law index $p$, (v) and the late-time afterglow slope respectively.
}
\label{tab:fit_params1}
\tablehead{
\colhead{GRB} & \colhead{Best} & \colhead{AICc} & \colhead{$t_{\rm rise}$} & \colhead{$t_{\rm decay}$} & \colhead{$\rm log_{10}(T_{\rm 1})$} & \colhead{$\rm log_{10}(T_{\rm 2}=T_{\rm a})$} & \colhead{$\rm log_{10}(T_{\rm 3})$} & \colhead{p} & \colhead{Late afterglow slope} \\
 \colhead{name} &  \colhead{model} & \colhead{} & \colhead{(s)}  & \colhead{(s)} & \colhead{(s)} & \colhead{(s)}  & \colhead{(s)} & \colhead{} &   \colhead{}
}
\startdata
221110A & G4 & 65 & ... & ... &  $1.96^{+0.04}_{-0.03}$ & ... & ... & $2.24^{+0.02}_{-0.02}$ & $-0.93^{+0.76}_{-0.74}$ \\
220521A & C2 & 47 & ... & ... &  $2.15^{+0.02}_{-0.001}$ & $2.84^{+0.05}_{-0.12}$& $3.40^{+0.6}_{-0.13}$ & $2.0^{+0.25}_{-0.04}$ & $-1.0^{+0.48}_{-0.48}$ \\
220117A & C4 & 251 & $54^{+4.9}_{-6.7}$ & $380^{+236}_{-150}$ &  $2.49^{+0.06}_{-0.0004}$ & $4.01^{+0.07}_{-0.15}$ & ... & $2.75^{+0.08}_{-0.19}$ & $-1.56^{+0.12}_{-0.6}$ \\
&  & &  $583^{+51}_{-67}$ & $4615^{+1016}_{-1432}$ & ... & ... & ... & ... & ... \\
220101A & C5 & 801 & $62^{+7.0}_{-9.6}$ & $235^{+61}_{-66}$ &  $2.21^{+0.01}_{-0.01}$ & $3.22^{+0.07}_{-0.13}$ & $4.90^{+0.16}_{-0.15}$ & $2.11^{+0.06}_{-0.03}$ & $-1.08^{+0.18}_{-0.52}$ \\
211024B & C4 &  442   & $1834^{+438}_{-378}$ & $1.2 \times 10^{4}{}^{+2.0\times10^{4}}_{-6066}$ &  $3.82^{+0.11}_{-0.04}$ & $5.0^{+0.12}_{-0.11}$ & ... & $2.01^{+0.04}_{-0.0003}$ & $-1.0^{+0.19}_{-0.5}$ \\
 & & & $5277^{+138}_{-239}$ & $1.0\times 10^{5}{}^{+1.6\times10^{4}}_{-2.5\times10^{4}}$ & ... & ... & ... & ... & ... \\
210905A & G3 & 272  & $22^{+3.0}_{-4.7}$ & $132^{+88}_{-64}$ &  ... & ... & ... & $2.14^{+0.01}_{-0.02}$  & $-0.86^{+0.76}_{-0.74}$ \\
 & & &  $27^{+1.8}_{-1.5}$  & $114^{+16}_{-15}$ & ... & ... & ... & ... & ... \\
210822A & H10 & 946 & ... & ... &   $1.83^{+0.01}_{-0.007}$ & ... & $3.90^{+0.03}_{-0.09}$  & $2.08^{+0.01}_{-0.007}$ & $-1.06^{+0.51}_{-0.50}$ \\
210731A & C1 & 46 & ... & ... &  $2.9^{+0.05}_{-0.07}$ & $4.43^{+0.19}_{-0.11}$ & ... & $2.57^{+0.30}_{-0.14}$ & $-1.42^{+0.58}_{-0.58}$ \\
210722A & G9 & 278  & $45^{+0.8}_{-0.9}$ & $1165^{+54}_{-101}$ &  ... & ... & $3.91^{+0.07}_{-0.03}$ & $2.0^{+0.05}_{-0.004}$ & $-0.75^{+0.77}_{-0.75}$ \\
& & &  $4383^{+390}_{-561}$  & $3.3\times 10^{4}{}^{+8695}_{-1.2\times10^{4}}$ & ... & ... & ... & ... & ... \\
210702A & H8 & 523 & $1983^{+596}_{-766}$ & $8537^{+1.3\times10^{4}}_{-4990}$ &  ... & ... & $2.60^{+0.98}_{-0.08}$ & $2.0^{+0.04}_{-0.005}$ & $-1.0^{+0.52}_{-0.50}$  \\
210619B & G10 & 1401 & ... & ... &  $2.33^{+0.02}_{-0.002}$ & ... & $4.08^{+0.03}_{-0.04}$ & $2.25^{+0.01}_{-0.007}$ & $-0.94^{+0.75}_{-0.75}$ \\
210610B & C5 &  651 & $22^{+0.7}_{-0.6}$ & $58^{+2.4}_{-2.0}$ &  $2.0^{+0.0001}_{-0.02}$ & $3.07^{+0.15}_{-0.02}$ & $4.84^{+0.16}_{-0.14}$ & $2.0^{+0.04}_{-0.002}$ &  $-1.0^{+0.2}_{-0.5}$ \\
210517A  & G1 & 20 & ... & ... &  ... & ... & ... & $2.2^{+0.05}_{-0.04}$ & $-0.92^{+0.77}_{-0.73}$ \\
210504A &  E3 & 104 & $49^{+34}_{-24}$ & $161^{+567}_{-110}$ &  $2.98^{+0.02}_{-0.03}$ & $4.10^{+0.86}_{-0.05}$ & ... & $2.03^{+0.20}_{-0.003}$ & $-1.02^{+0.5}_{-0.5}$ \\
210420B &  E3 & 121 & $106^{+39}_{-31}$ & $329^{+331}_{-148}$ &  $3.13^{+0.02}_{-0.27}$ & $4.04^{+0.58}_{-0.27}$  & ... & $2.50^{+0.42}_{-0.06}$ & $-1.38^{+0.53}_{-0.53}$ \\
210411C &  G9 & 70 &  $27^{+23}_{-9}$ & $76^{+283}_{-32}$ &  ... & ... & $3.98^{+0.004}_{-0.17}$ & $2.04^{+0.02}_{-0.03}$ & $-0.78^{+0.76}_{-0.74}$ \\
 & & &  $262^{+94}_{-92}$  & $1304^{+1566}_{-768}$ & ... & ... & ... & ... & ... \\
210210A  & C1 & 89 & ... & ... &  $1.89^{+0.02}_{-0.02}$ & $3.53^{+0.07}_{-0.08}$ & ... & $2.66^{+0.06}_{-0.06}$ & $-1.49^{+0.16}_{-0.53}$ \\
201104B &   C3 & 112  & $30^{+11}_{-15}$ & $272^{+902}_{-227}$ &  $2.02^{+0.02}_{-0.02}$ & $2.52^{+0.05}_{-0.02}$ & ... & $2.23^{+0.05}_{-0.01}$ & $-1.17^{+0.19}_{-0.5}$ \\
201024A & G1 & 36 & ... & ... &  ... & ... & ... & $2.27^{+0.04}_{-0.03}$ & $-0.96^{+0.76}_{-0.74}$ \\
201021C & H4 & 36 & ... & ... &  $2.22^{+0.04}_{-0.05}$ & ... & ... & $2.16^{+0.04}_{-0.05}$ & $-1.12^{+0.52}_{-0.47}$ \\
201020A & C1 & 47 & ... & ... &  $2.46^{+0.01}_{-0.06}$ & $3.93^{+0.10}_{-0.11}$ &  ... & $2.76^{+0.09}_{-0.12}$ & $-1.57^{+0.10}_{-0.56}$ \\
201014A  & E1 & 30 & ... & ... & $2.50^{+0.06}_{-0.03}$ & $5.25^{+0.14}_{-0.22}$ & ... & $2.23^{+0.06}_{-0.10}$ &  $-1.17^{+0.16}_{-0.55}$ \\
200205B &  C6 & 523 & $125^{+4.4}_{-2.5}$  & $1649^{+344}_{-220}$ &  $2.98^{+0.01}_{-0.01}$ & $4.24^{+0.07}_{-0.03}$ & $5.57^{+1.22}_{-0.03}$ & $2.68^{+0.13}_{-0.06}$ & $-1.51^{+0.54}_{-0.54}$ \\
& & &  $36^{+2.9}_{-2.9}$  & $147^{+25}_{-23}$ & ... & ... & ... & ... & ... \\
191221B &  C5 &  954 & $68^{+1.45}_{-2.21}$ & $1315^{+139}_{-241}$ & $2.20^{+0.01}_{-0.01}$ & $3.0^{+0.01}_{-0.01}$ & $5.0^{+0.04}_{-0.12}$ & $2.36^{+0.02}_{-0.01}$ & $-1.27^{+0.23}_{-0.5}$ \\
191011A & C1 & 39 & ... & ... &  $2.04^{+0.05}_{-0.01}$ & $2.94^{+0.05}_{-0.12}$ & ... & $2.45^{+0.03}_{-0.10}$  & $-1.34^{+0.22}_{-0.56}$ \\
191004B & H4 & 73 & ... & ... &  $2.01^{+0.02}_{-0.04}$ & ... & ... & $2.09^{+0.02}_{-0.02}$  & $-1.07^{+0.51}_{-0.49}$ \\
190829A &  C5 & 789 & $386^{+46}_{-32}$ & $1074^{+234}_{-123}$ & $2.24^{+0.001}_{-0.04}$ & $3.71^{+0.05}_{-0.05}$ & $5.6^{+0.22}_{-0.08}$ & $2.09^{+0.02}_{-0.02}$ & $-1.07^{+0.23}_{-0.51}$  \\
190719C &  C4 & 685 & $45^{+6}_{-5}$ & $221^{+162}_{-75}$ &  $2.87^{+0.02}_{-0.04}$ & $4.81^{+0.08}_{-0.07}$ & ... & $2.61^{+0.10}_{-0.06}$ &$-1.46^{+0.08}_{-0.53}$  \\
 & & & $30^{+7.7}_{-6.4}$  & $173^{+279}_{-90}$ & ... & ... & ... & ... & ... \\
190114A & E6 & 130 & $13^{+8.4}_{-2.8}$ & $27^{+29}_{-7}$ &  $2.62^{+0.01}_{-0.02}$ & $3.86^{+0.67}_{-0.03}$ & $4.61^{+1.83}_{-0.05}$ & $2.0^{+0.59}_{-0.02}$ & $-1.0^{+0.49}_{-0.49}$ \\
 & & &  $606^{+591}_{-443}$  & $4084^{+5800}_{-3612}$ & ... & ... & ... & ... & ... \\
190106A &   C6 & 290 & $20^{+1.3}_{-2.1}$ & $770^{+350}_{-366}$ &  $2.44^{+0.03}_{-0.01}$ & $3.73^{+0.10}_{-0.06}$ & $4.91^{+0.13}_{-0.14}$ &  $2.08^{+0.06}_{-0.05}$ & $-1.06^{+0.17}_{-0.53}$ \\
 & & &  $29^{+30}_{-16}$  & $110^{+251}_{-78}$ & ... & ... & ... & ... & ... \\
181110A &   C6  & 588 &  $50^{+1.5}_{-1.5}$  & $934^{+36}_{-66}$ &  $2.49^{+0.01}_{-0.01}$ & $3.20^{+0.03}_{-0.01}$ & $3.74^{+0.03}_{-0.03}$ & $2.0^{+0.03}_{-0.002}$ & $-1.0^{+0.21}_{-0.50}$ \\
 & & & $66^{+1.2}_{-1.4}$  & $1368^{+120}_{-182}$ & ... & ... & ... & ... & ... \\
181020A & G12 & 1669  & $91^{+0.78}_{-0.92}$ & $2240^{+64}_{-122}$ &  $1.80^{+0.01}_{-0.01}$ & ... & $4.22^{+0.03}_{-0.01}$ & $2.23^{+0.01}_{-0.01}$ & $-0.92^{+0.75}_{-0.75}$ \\
 & & &  $64^{+4.4}_{-7.0}$  & $2559^{+1560}_{-1305}$ & ... & ... & ... & ... & ... \\
181010A &  C5 & 208 & $22^{+20}_{-11}$ & $73^{+219}_{-42}$ & $2.30^{+0.11}_{-0.26}$ & $3.12^{+0.2}_{-0.27}$ &  $5.11^{+0.37}_{-0.28}$ & $2.01^{+0.07}_{-0.01}$ & $-1.0^{+0.15}_{-0.49}$ \\
180728A &  E4 & 978 & $592^{+144}_{-136}$ & $2335^{+2820}_{-1094}$ &  $3.32^{+0.01}_{-0.01}$ & $4.42^{+0.04}_{-0.01}$ & ... & $2.08^{+0.02}_{-0.01}$ & $-1.06^{+0.23}_{-0.51}$ \\
 & & & $711^{+956}_{-458}$ & $2068^{+1.7\times10^{4}}_{-1475}$ & ... & ... & ... & ... & ... \\
180720B &  C4 & 3434 & $165^{+3.2}_{-4.2}$ & $2510^{+175}_{-311}$ &  $2.52^{+0.004}_{-0.02}$ & $3.53^{+0.02}_{-0.02}$ & ... & $2.48^{+0.01}_{-0.01}$ & $-1.36^{+0.24}_{-0.51}$ \\
 & & &   $14.11^{+1.18}_{-1.04}$  & $55^{+17}_{-12}$ & ... & ... & ... & ... & ... \\
180624A &  E4 & 1487  &  $22^{+1.01}_{-0.71}$ & $64^{+10}_{-6}$ &  $2.75^{+0.002}_{-0.01}$ & $4.59^{+0.04}_{-0.19}$ & ... & $2.0^{+0.04}_{-0.004}$ & $-1.0^{+0.2}_{-0.5}$ \\
 & & & $98^{+6.8}_{-8.4}$  & $305^{+54}_{-57}$ & ... & ... & ... & ... & ... \\
180620B &  E6 & 367 & $76^{+5.1}_{-6.0}$ & $840^{+203}_{-274}$ &  $2.64^{+0.02}_{-0.03}$ & $5.09^{+0.07}_{-0.05}$ & $5.68^{+0.09}_{-0.08}$  & $2.0^{+0.03}_{-0.0002}$ & $-1.0^{+0.21}_{-0.5}$ \\
 & & &  $2155^{+905}_{-806}$  & $8864^{+1.3\times10^{4}}_{-4540}$ & ... & ... & ... & ... & ... \\
\enddata
\end{deluxetable}

\begin{deluxetable}{cccccccccc}
\tabletypesize{\scriptsize}
\renewcommand{\thetable}{}
\makeatletter
\renewcommand{\@makecaption}[2]{#2} 
\makeatother
\caption{\textbf{Table \ref{tab:fit_params1}} (Continued)}
\label{tab:fit_params2}
\tablehead{
\colhead{GRB} & \colhead{Best} & \colhead{AICc}  & \colhead{$t_{\rm rise}$} & \colhead{$t_{\rm decay}$}  &\colhead{$\rm log_{10}(T_{\rm 1})$} & \colhead{$\rm log_{10}(T_{\rm 2}=T_{\rm a})$} & \colhead{$\rm log_{10}(T_{\rm 3})$} & \colhead{p} &  \colhead{Late afterglow slope} \\
 \colhead{name} &  \colhead{model} & \colhead{} & \colhead{(s)}  & \colhead{(s)} &  \colhead{(s)} & \colhead{(s)}  & \colhead{(s)} & \colhead{} & \colhead{} 
}
\startdata
180510B  & H1 & 67 & ... & ... &   ... & ...  & ... & $2.30^{+0.03}_{-0.03}$ & $-1.22^{+0.52}_{-0.48}$ \\
180404A  & E1 & 22 & ... & ... &  $2.24^{+0.04}_{-0.04}$ & $4.44^{+0.32}_{-0.15}$ & ... & $2.18^{+0.15}_{-0.11}$ & $-1.13^{+0.56}_{-0.56}$ \\
180329B &  E6 & 440 &  $33^{+3.1}_{-2.7}$ & $117^{+36}_{-23}$ &  $2.23^{+0.01}_{-0.01}$ & $3.89^{+0.04}_{-0.04}$ & $4.39^{+0.82}_{-0.10}$ & $2.01^{+0.08}_{-0.004}$ & $-1.0^{+0.14}_{-0.50}$ \\
 & & & $38^{+3.9}_{-3.6}$  & $164^{+54}_{-38}$ &  ... & ... & ... & ... & ... \\
180325A & C3 & 381 &  $4.92^{+0.99}_{-0.54}$ & $10^{+2.1}_{-0.8}$ &  $1.82^{+0.01}_{-0.0003}$ & $3.44^{+0.01}_{-0.03}$ & ... & $3.0^{+0.001}_{-0.02}$ & $-1.75^{+0.25}_{-0.51}$ \\
180205A &  G8 & 89 &  $21^{+12}_{-7}$ & $65^{+149}_{-36}$ &  ... & ... & $3.36^{+0.39}_{-0.05}$ & $2.01^{+0.18}_{-0.03}$ & $-0.76^{+0.81}_{-0.76}$ \\
171222A &  C3 & 212 &  $120^{+14}_{-13}$ & $437^{+118}_{-81}$ &  $3.07^{+0.03}_{-0.02}$ & $5.38^{+0.21}_{-0.07}$ & ... & $2.49^{+0.23}_{-0.01}$ & $-1.37^{+0.51}_{-0.51}$ \\
171205A  & C1 & 333 & ... & ... &  $3.89^{+0.02}_{-0.03}$ & $5.02^{+0.1}_{-0.04}$ & ... & $2.01^{+0.11}_{-0.01}$ & $-1.01^{+0.10}_{-0.50}$ \\
171020A  & G4 & 33 & ... & ... &  $2.72^{+0.05}_{-0.09}$ & ...  & ... & $2.18^{+0.04}_{-0.05}$ & $-0.89^{+0.76}_{-0.73}$ \\
170714A &  C4 & 3227 & $2286^{+170}_{-276}$ & $2.6\times10^{4}{}^{+9083}_{-1.1\times10^{4}}$ &   $2.58^{+0.001}_{-0.001}$ & $3.08^{+0.001}_{-0.001}$ & ... & $2.97^{+0.02}_{-0.01}$ & $-1.73^{+0.23}_{-0.51}$ \\
 & & &  $5094^{+88}_{-118}$  & $1.1\times10^{5}{}^{+7687}_{-1.5\times10^{4}}$ & ... & ... & ... & ... & ... \\
170705A &  C5 & 992 &  $96^{+0.60}_{-0.64}$ & $1704^{+19}_{-39}$ &   $2.40^{+0.001}_{-0.01}$ & $3.88^{+0.03}_{-0.01}$ & $5.36^{+0.32}_{-0.06}$  & $2.0^{+0.02}_{-0.0003}$ & $-1.0^{+0.23}_{-0.5}$ \\
170607A &  E4 & 576 & $41^{+6.1}_{-4.3}$ & $122^{+18}_{-9}$ & $2.88^{+0.02}_{-0.02}$ & $5.12^{+0.07}_{-0.07}$ & ... & $2.03^{+0.06}_{-0.02}$ & $-1.02^{+0.17}_{-0.51}$ \\
 & & &  $4337^{+609}_{-791}$  & $3.0\times 10^{4}{}^{+1.4\times10^{4}}_{-1.4\times10^{4}}$ & ... & ... & ... & ... & ... \\
170604A &  H3 & 535 & $8.8^{+0.55}_{-0.42}$  & $33^{+2.8}_{-3.1}$ &  ... & ... & ... & $2.01^{+0.02}_{-0.01}$ &  $-1.01^{+0.51}_{-0.50}$ \\
 & & &  $108^{+5.2}_{-11}$   & $2515^{+1032}_{-1181}$ & ... & ... & ... & ... & ... \\
170531B &  G3 & 397 & $40^{+2.0}_{-2.4}$ & $215^{+38}_{-37}$ &  ... & ... & ... & $2.33^{+0.07}_{-0.02}$ & $-1.0^{+0.78}_{-0.76}$ \\
 & & & $90^{+11}_{-10}$  & $621^{+283}_{-178}$ &  ... & ... & ... & ... & ... \\
170519A &  E4 & 820 & $15^{+0.29}_{-0.28}$ & $53^{+2.8}_{-2.7}$ &  $2.64^{+0.01}_{-0.02}$ & $3.85^{+0.05}_{-0.03}$ & ... & $1.81^{+0.02}_{-0.01}$ & $-0.86^{+0.2}_{-0.5}$ \\
 & & &  $436^{+473}_{-241}$  & $1643^{+2055}_{-794}$ & ... & ... & ... & ... & ... \\
170405A &  H9 & 420 &  $22.10^{+4.65}_{-3.73}$ & $109^{+91}_{-41}$ &   ... & ... & $2.21^{+0.18}_{-0.03}$ & $2.06^{+0.54}_{-0.08}$ & $-1.05^{+0.74}_{-0.54}$ \\
 & & &  $2661^{+1102}_{-1169}$  & $1.3\times 10^{4}{}^{+1.6\times10^{4}}_{-8280}$ & ... & ... & ... & ... & ... \\
170202A &  C3 & 108 & $10^{+4.4}_{-2.7}$ & $31^{+37}_{-13}$ & $2.47^{+0.02}_{-0.03}$ & $3.40^{+0.06}_{-0.09}$ & ... & $2.18^{+0.03}_{-0.09}$ & $-1.14^{+0.21}_{-0.55}$ \\
170113A &  C4 & 273 & $36^{+1.6}_{-1.9}$ & $546^{+161}_{-141}$ & $2.18^{+0.01}_{-0.06}$ & $3.44^{+0.07}_{-0.06}$ & ... & $2.31^{+0.03}_{-0.01}$ & $-1.23^{+0.21}_{-0.51}$  \\
 & & & $5.89^{+2.03}_{-1.70}$ & $15^{+12}_{-7}$ & ... & ... & ... & ... & ... \\
161219B &  C6 & 737 & $70^{+3.6}_{-3.7}$ & $245^{+53}_{-43}$ & $2.72^{+0.02}_{-0.02}$ &  $4.78^{+0.02}_{-0.03}$ & $6.39^{+0.83}_{-0.06}$ & $2.0^{+0.004}_{-0.001}$ & $-1.0^{+0.3}_{-0.5}$  \\
 & & &  $1247^{+63}_{-75}$  & $1.3\times 10^{4}{}^{+1591}_{-2742}$ & ... & ... & ... & ... & ... \\
161117A & G12 & 560 & $35^{+2.5}_{-2.3}$ & $134^{+28}_{-22}$ &  $2.65^{+0.002}_{-0.04}$ & ... & $4.99^{+0.26}_{-0.14}$ &  $2.02^{+0.02}_{-0.02}$ & $-0.77^{+0.76}_{-0.74}$ \\
 & & & $3703^{+507}_{-696}$  & $2.3\times 10^{4}{}^{+1.2\times10^{4}}_{-1.1\times10^{4}}$ & ... & ... & ... & ... & ... \\
161108A &  G6 & 416 & $55^{+4.6}_{-5.3}$ & $253^{+78}_{-67}$ &  $2.73^{+0.02}_{-0.05}$ & ... & ... & $2.0^{+0.007}_{-0.0004}$ & $-0.75^{+0.75}_{-0.75}$ \\
 & & &  $83^{+150}_{-52}$   & $330^{+2886}_{-257}$ & ... & ... & ... & ... & ... \\
161017A &  C6 & 1562 &  $86^{+1.1}_{-1.1}$  & $1626^{+35}_{-64}$ & $1.78^{+0.01}_{-0.02}$ & $2.34^{+0.03}_{-0.02}$ & $4.80^{+0.09}_{-0.09}$ & $2.36^{+0.01}_{-0.01}$ & $-1.27^{+0.23}_{-0.50}$ \\
& & & $108^{+2.0}_{-2.3}$  & $4521^{+294}_{-556}$ &  ... & ... & ... & ... & ... \\
161014A   & C2 & 68 & ... & ... & $2.05^{+0.02}_{-0.001}$ &  $3.26^{+0.06}_{-0.08}$ & $4.26^{+0.08}_{-0.11}$ & $2.83^{+0.10}_{-0.09}$ & $-1.62^{+0.08}_{-0.55}$ \\
160804A & G12 & 1464 &  $164^{+3.0}_{-2.9}$ & $4030^{+88}_{-152}$ &  $3.15^{+0.01}_{-0.002}$ & ... &  $5.47^{+0.23}_{-0.02}$ & $2.02^{+0.03}_{-0.01}$ & $-0.76^{+0.76}_{-0.75}$ \\
 & & &  $4843^{+2294}_{-1950}$  & $1.8\times 10^{4}{}^{+3.5\times10^{4}}_{-8367}$ & ... & ... & ... & ... & ... \\
160425A &  G5 & 1004 & $55^{+2.6}_{-2.4}$ & $232^{+35}_{-29}$ &  $2.71^{+0.01}_{-0.03}$ & ... & ... & $2.17^{+0.02}_{-0.02}$ &  $-0.88^{+0.76}_{-0.74}$ \\
160410AS &  E3 & 77 & $82^{+28}_{-42}$ & $516^{+457}_{-376}$ &  $2.37^{+0.03}_{-0.05}$ & $3.88^{+0.44}_{-0.53}$  & ... & $2.30^{+0.56}_{-0.05}$ & $-1.22^{+0.53}_{-0.53}$ \\
160228A &  H8 & 95 &  $11^{+42}_{-7}$ & $63^{+495}_{-53}$ &  ... & ... & $2.08^{+0.05}_{-0.02}$ & $2.03^{+0.35}_{-0.03}$ & $-1.03^{+0.67}_{-0.52}$ \\
160227A &  E4 & 859 & $20^{+0.7}_{-0.7}$ &  $63^{+6.2}_{-5.8}$ & $2.53^{+0.01}_{-0.01}$ & $4.78^{+0.05}_{-0.02}$ & ... & $2.01^{+0.02}_{-0.01}$ & $-1.01^{+0.22}_{-0.50}$ \\
 & & &  $43^{+1.03}_{-0.97}$  & $302^{+16}_{-15}$ & ... & ... & ... & ... & ... \\
160203A & C1 & 24 & ... & ... & $2.40^{+0.30}_{-0.03}$ &  $3.13^{+0.4}_{-0.1}$ & ... & $2.23^{+0.11}_{-0.11}$ & $-1.17^{+0.04}_{-0.56}$ \\
160131A &  E4 &  582 & $102^{+8.5}_{-13.7}$ & $769^{+196}_{-254}$ &  $2.25^{+0.001}_{-0.02}$ & $4.95^{+0.05}_{-0.03}$ & ... &  $2.83^{+0.12}_{-0.01}$ & $-1.62^{+0.51}_{-0.51}$ \\
 & & & $2800^{+355}_{-543}$  & $2.7\times 10^{4}{}^{+1.8\times10^{4}}_{-1.5\times10^{4}}$ & ... & ... & ... & ... & ... \\
160121A & C1  & 44 & ... & ... &   $2.33^{+0.04}_{-0.05}$ & $4.19^{+0.11}_{-0.10}$ & ... & $2.59^{+0.10}_{-0.10}$ & $-1.44^{+0.08}_{-0.55}$ \\
151215A & C1 & 25 & ... & ... &  $2.50^{+0.003}_{-0.09}$ & $3.03^{+0.17}_{-0.09}$ & ... & $2.14^{+0.13}_{-0.04}$ & $-1.11^{+0.52}_{-0.52}$ \\
151031A & C1 & 45 & ... & ... &  $2.77^{+0.04}_{-0.02}$ & $3.51^{+0.11}_{-0.04}$ & ... & $2.0^{+0.16}_{-0.02}$ & $-1.0^{+0.49}_{-0.49}$ \\
151029A & H1 & 15 & ... & ... &  ... & ... & ... & $2.17^{+0.06}_{-0.06}$ & $-1.13^{+0.53}_{-0.46}$ \\
151027B &  C3 & 66 &  $730^{+696}_{-577}$ &  $2541^{+1.4\times10^{4}}_{-1969}$ & $2.73^{+0.04}_{-0.03}$ & $4.32^{+0.32}_{-0.03}$ & ... & $2.23^{+0.35}_{-0.07}$ &  $-1.17^{+0.54}_{-0.54}$  \\
151027A &  C6 & 909 & $48^{+0.7}_{-1.3}$ & $1133^{+121}_{-195}$ &  $2.19^{+0.01}_{-0.004}$ & $3.39^{+0.03}_{-0.01}$ & $4.03^{+0.03}_{-0.06}$ & $2.01^{+0.03}_{-0.002}$ & $-1.0^{+0.21}_{-0.50}$ \\
 & & &  $103^{+1.92}_{-2.34}$  & $2935^{+204}_{-378}$ & ... & ... & ... & ... & ... \\
151021A &  C4 & 294 &  $46^{+1.60}_{-2.16}$ & $1000^{+166}_{-240}$ &  $2.34^{+0.02}_{-0.004}$ & $3.31^{+0.03}_{-0.05}$ & ... & $2.50^{+0.03}_{-0.04}$ & $-1.37^{+0.22}_{-0.52}$ \\
 & & & $45^{+19}_{-13}$  & $139^{+305}_{-62}$ & ... & ... & ... & ... & ... \\
150821A &  G6 & 670 & $167^{+20}_{-15}$ & $466^{+134}_{-73}$ &   $3.15^{+0.16}_{-0.003}$ & ... & ... & $2.86^{+0.01}_{-0.15}$ & $-1.40^{+0.74}_{-0.69}$ \\
 & & &  $54^{+25}_{-21}$  & $210^{+317}_{-128}$ & ... & ... & ... & ... & ... \\
150818A &  E3 &  301  &  $10^{+1.3}_{-2.4}$ & $411^{+434}_{-288}$ &  $3.82^{+0.02}_{-0.07}$ & $6.82^{+0.02}_{-0.08}$ & ... & $2.0^{+0.08}_{-0.01}$ & $-1.0^{+0.13}_{-0.49}$ \\
150727A &  E3 & 473 & $429^{+41}_{-57}$ & $3579^{+674}_{-1064}$ &  $3.70^{+0.03}_{-0.07}$ & $6.10^{+0.4}_{-0.52}$ & ... & $2.14^{+0.20}_{-0.08}$ & $-1.11^{+0.54}_{-0.54}$ \\
150323A & C1 & 1669 & ... & ... &  $2.87^{+0.01}_{-0.01}$ & $4.0^{+0.06}_{-0.05}$ & ... & $2.31^{+0.08}_{-0.10}$  & $-1.23^{+0.13}_{-0.55}$ \\
\enddata
\end{deluxetable}

\begin{deluxetable}{cccccccccc}
\tabletypesize{\scriptsize}
\renewcommand{\thetable}{}
\makeatletter
\renewcommand{\@makecaption}[2]{#2} 
\makeatother
\caption{\textbf{Table \ref{tab:fit_params1}} (Continued)}
\label{tab:fit_params3}
\tablehead{
\colhead{GRB} & \colhead{Best} & \colhead{AICc}  & \colhead{$t_{\rm rise}$} & \colhead{$t_{\rm decay}$} & \colhead{$\rm log_{10}(T_{\rm 1})$} & \colhead{$\rm log_{10}(T_{\rm 2}=T_{\rm a})$} & \colhead{$\rm log_{10}(T_{\rm 3})$} & \colhead{p} & \colhead{Late afterglow slope} \\
 \colhead{name} &  \colhead{model} & \colhead{} & \colhead{(s)}  & \colhead{(s)} & \colhead{(s)} & \colhead{(s)}  & \colhead{(s)} & \colhead{} &  \colhead{} 
}
\startdata
150314A &  G11 & 704 &  $4360^{+4342}_{-1205}$ & $6.0\times 10^{4}{}^{+3.9\times10^{6}}_{-3.4\times10^{4}}$ &  $1.98^{+0.01}_{-0.002}$ & ... & $3.22^{+0.02}_{-0.21}$ & $2.21^{+0.02}_{-0.01}$ & $-0.91^{+0.76}_{-0.75}$  \\
150301B & G1 & 45 & ... & ... &  ... & ... & ... & $2.46^{+0.03}_{-0.03}$ & $-1.09^{+0.76}_{-0.74}$ \\
150206A &  H3 & 813 & $652^{+55}_{-106}$ &  $1.1\times 10^{4}{}^{+5521}_{-5531}$ &  ... & ... & ... & $2.43^{+0.01}_{-0.01}$ & $-1.32^{+0.5}_{-0.5}$ \\
 & & &  $426^{+24}_{-37}$  & $1.9\times 10^{4}{}^{+8544}_{-8716}$ & ... & ... & ... & ... & ... \\
141225A & G4 & 15 & ... & ... &  $3.34^{+0.11}_{-0.14}$ & ... & ... & $2.31^{+0.22}_{-0.17}$ & $-0.99^{+0.83}_{-0.68}$ \\
141221A &  H3 & 69 &  $126^{+51}_{-35}$ & $462^{+698}_{-194}$ &  ... & ... & ... & $2.01^{+0.02}_{-0.01}$ & $-1.01^{+0.51}_{-0.49}$ \\
 & & &  $53^{+26}_{-31}$  & $372^{+1362}_{-320}$ & ... & ... & ... & ... & ... \\
141220A & C1 & 34 & ... & ... &  $1.96^{+0.02}_{-0.002}$ & $2.46^{+0.10}_{-0.01}$ & ... & $2.52^{+0.09}_{-0.001}$ & $-1.39^{+0.1}_{-0.5}$ \\
141121A &  C4 &  882 &  $53^{+31}_{-23}$ & $189^{+307}_{-123}$ & $3.12^{+0.06}_{-0.01}$ & $5.47^{+0.06}_{-0.13}$ & ... &  $2.99^{+0.03}_{-0.23}$ & $-1.75^{+0.28}_{-0.62}$ \\
 & & &  $2756^{+314}_{-625}$  & $2.3\times 10^{4}{}^{+1.6\times10^{4}}_{-1.4\times10^{4}}$ & ... & ... & ... & ... & ... \\
141004A & G7 & 28 & ... & ... &  ... & ... & $3.62^{+0.08}_{-0.06}$ & $2.0^{+0.10}_{-0.01}$ & $-0.75^{+0.79}_{-0.75}$ \\
140907A &  C3 & 111 &  $66^{+27}_{-27}$ & $530^{+749}_{-405}$ & $2.37^{+0.04}_{-0.11}$ & $3.37^{+0.06}_{-0.29}$ & ... &  $2.0^{+0.04}_{-0.002}$  & $-1.0^{+0.2}_{-0.5}$  \\
140710A &  G2 & 47 &  $36^{+27}_{-7}$ & $105^{+110}_{-31}$ &  ... & ... & ... & $2.18^{+0.06}_{-0.04}$ & $-0.89^{+0.77}_{-0.74}$ \\
140703A &  C5 & 176 &  $4.35^{+4.25}_{-1.03}$ & $8.7^{+17.3}_{-2.1}$ & $2.41^{+0.02}_{-0.01}$ & $3.96^{+0.03}_{-0.05}$ & $4.65^{+0.05}_{-0.07}$ & $2.85^{+0.08}_{-0.10}$ &  $-1.64^{+0.13}_{-0.56}$ \\
140629A &  C5 & 119 &  $54^{+42}_{-23}$ & $172^{+342}_{-77}$ & $2.33^{+0.02}_{-0.14}$ & $3.23^{+0.15}_{-0.15}$ & $4.42^{+0.16}_{-0.21}$ & $2.33^{+0.10}_{-0.15}$ & $-1.24^{+0.08}_{-0.58}$ \\
\enddata
\end{deluxetable}

\section{Sample of GRBs and the prompt properties}
\label{sec:sample_prompt_properties}

\subsection{Prompt isotropic energy of spectrum}
\label{app:Eiso_computation}

To compute the isotropic-equivalent energy ($E_{\rm iso}$) of the prompt emission in GRBs, it is essential to accurately model the spectrum and integrate it over a specified energy range. This calculation involves accounting for the spectral shape, which may be best described by different models: the Band model \citep{band+93}, simple broken power-law (SBPL) model, an exponential cut-off power-law (CPL) model or a simple power-law model (PL), depending on the data source (Fermi-GBM or Swift-BAT) and energy range. Additionally, a k-correction must be applied to account for distance and redshift effects in cosmological sources.

The Swift-BAT observations cover a relatively narrow energy range (15–150 keV), so the spectrum often best-fit with a simple PL model. For $E_{\rm iso}$ calculations using this model, we followed the same method presented in \citet{Tang+19, Xu+21}. 
The isotropic energy of the prompt emission is 
\begin{equation}
\centering
    E^{'}_{\gamma, \rm iso} = \frac{4 \pi {d_{\rm L}^2 S}}{(1+z)},
\end{equation}
where $S$ is the fluence in the BAT energy band, and $d_{\rm L}$ is the luminosity distance. A k-correction is then applied such that  $E_{\rm iso} = k E^{'}_{\gamma, \rm iso}$, where $k$ is the k-correction factor defined as 
\begin{equation}
\centering
k = \frac{\int_{e_1/(1+z)}^{e_2/(1+z)} E \phi (E) dE}{\int_{e_1}^{e_2}E \phi (E) dE}.
\end{equation}
To calculate the k-correction, a simple power-law spectral model is adopted, i.e.
$\phi (E) \propto E^{\Gamma_{\rm BAT}}$, where $\phi(E)$ is the source
photon spectrum and $\Gamma_{\rm BAT}$ is the photon spectral index.
The estimate on the isotropic energy in the BAT band, $E_{\rm iso}$ is
thus computed as
\begin{equation}
\centering
    E_{\rm iso} = \frac{4 \pi {d_{\rm L}^2 S}}{(1+{z})^{3-{\Gamma_{\rm BAT}}}}.
\end{equation}

For the cases where the spectrum is best fitted with the CPL or Band model, based on data from either Swift-BAT and Fermi-GBM, we compute the $E_{\rm iso}$ following the method presented in \citet{ZGA14}.

The Swift-BAT best-fit parameters (Fluence, $S$, photon spectral index $\Gamma_{\rm BAT}$ and peak energy $\rm E_{\rm pk}$ for cutoff power-law model) are taken from the BAT burst catalog\footnote{\url{https://swift.gsfc.nasa.gov/results/batgrbcat}} \citet{Gehrels+04}. 
The Fermi-GBM best-fit parameters (Fluence, low-energy power-law spectral index, and peak energy $\rm E_{pk}$) are obtained from the Fermi-GBM burst catalog available at HEASARC\footnote{\url{https://heasarc.gsfc.nasa.gov/W3Browse/fermi/fermigbrst.html}} \citep{vonKienlin2020} \footnote{Note that some GRBs observed between June 2008 and January 2022 by Fermi-GBM do not have fitted parameters in this catalog.}. 
Finally, the luminosity distance ${d_{\rm L}(z)}$ is computed assuming a flat Lambda cold dark matter $\Lambda$CDM
cosmological model with parameters $\Omega_m = 0.286$ and ${H_{\rm 0}} =
70 ~{\rm km~s^{-1}~Mpc^{-1}}$ \citep{Hinshaw+09}, with $z$ representing the redshift.  

\subsection{Prompt peak energy of spectrum}
\label{app:Epk_computation}

As mentioned in the previous subsection, the peak energy ($E_{\rm pk}$) is extracted directly from the Fermi-GBM and Swift-BAT catalogs when the spectrum is best fitted by either the Band model or the CPL model. However, in the Swift-BAT catalog, the best fit for the time-averaged spectral analysis is not always defined. Therefore, a criterion was set to restrict the reduced $\chi^2$ value to between 0.9 and 1.3 for all GRBs. Using this criterion, if the $E_{\rm pk}$ value is not properly constrained (appears as 9999.999 in the Swift-BAT catalog) or if both the high and low errors are significantly larger than $E_{\rm pk}$, those fit parameters were excluded from use. In this case, we used the fit parameters obtained with the power-law model. When the spectra in the catalogs were only best fitted by the power-law model or did not meet the CPL selection criteria mentioned above, $E_{\rm pk}$ was estimated using the correlation between peak energy and spectral index, derived from fitting a single power-law to large Fermi-GBM and Swift-BAT datasets, as parameterized by \citet{Virgili2012}. This method, commonly applied in the literature \citep{Zhang07, Racusin09}, provides results consistent with those derived from the Fermi-GBM data.


\subsection{Sample and prompt properties}
\label{app:prompt_properties}
\setcounter{table}{0} 
Table \ref{tab:prompt_parameters1}
provides the relevant parameters for our sample, including
(i) GRB name, (ii) redshift ($z$), (iii) Swift Trigger number (iv) flux conversion factor (from XRT light curves repository), (v) burst duration ($T_{\rm 90}$ from the BAT burst catalog), (vi) isotropic energy
(${E_{\rm iso}}$) and (vii) peak energy ($E_{\rm pk}$) in both the Fermi-GBM band ($8 \rm keV-40 \rm MeV$) and Swift-BAT band ($15 - 150 \rm keV$). Also included are (viii) the Instrument type and, in parenthesis the best-fit model type.  

From the table, using the burst duration $T_{90}$, the isotropic energy $E_{\rm iso}$, and the peak energy $E_{\rm pk}$ for all 89 GRBs in our sample, we find the average values to be $\left < \log_{10} T_{90}\right > = 1.71 \pm 0.06$, $\left < \log_{10} E_{\rm iso}\right > = 52.07 \pm 0.10$, and $\left < \log_{10} E_{\rm pk}\right > = 2.17 \pm 0.04$.
Additionally, we examine whether GRBs observed by Fermi-GBM behave differently from the full sample, and find that the average peak energy is of the same order of magnitude: $\left < \log_{10} E_{\rm pk}\right > = 2.24 \pm 0.05$.


\begin{deluxetable}{ccccccccccccc}[H]
\tabletypesize{\scriptsize}
\tablecaption{A sample of 89 GRBs used in this study. Column 1: GRB name, Column 2: Redshift, Column 3: Swift trigger number, Column 4: The flux conversion factor, $\mathrm{C_F}$, as defined in \citet[Sec.~2.2]{DPR22},  
which converts the XRT count rate (counts s$^{-1}$) to flux (erg cm$^{-2}$ s$^{-1}$),  
and is used to compute the isotropic-equivalent energy $E_{\rm iso,f}$ emitted in each flare.  
$\mathrm{C_F}$ values are taken from the \textit{Swift} repository\footnote{\url{https://www.swift.ac.uk/xrt_curves/}}. Column 5: Burst duration and its error, Column 6: Spectral index and its error, Column 7: Isotropic equivalent energy in log scale and its error, Column 8: Peak energy and its errors. Column 9: The instrument from which the data was obtained, the model types used to fit the data, and the resulting parameters used to compute the isotropic equivalent energy of each burst using different methods, as explained in Appendix \ref{sec:sample_prompt_properties}.}
\label{tab:prompt_parameters1} 
\tablehead{ 
\colhead{GRB} & \colhead{z} & \colhead{Trigger} & \colhead{ $\mathrm{C_F}$} & \colhead{$T_{90}$} & \colhead{Spectral Index}  & \colhead{$\log_{10}(E_{\rm iso})$}  & \colhead{$E_{\rm pk}$} &  \colhead{Instrument} \\
\colhead{} & \colhead{} & \colhead{} & \colhead{$\times10^{-11}$}  & \colhead{} & \colhead{} \\
 \colhead{}  & \colhead{} & \colhead{} & \colhead{($\rm erg ~ cm^{-2} $)} & \colhead{(s)} &\colhead{} & \colhead{(erg)} & \colhead{(keV)} &  \colhead{($\rm model ~type$)} 
}
\startdata  
221110A & 4.06 & 1136936 & 3.40 & 9.0$\pm$2.1 & $-1.41\pm0.16$ & 51.82$\pm$0.17 & 305$\pm$302 & Swift-BAT (PL fit)  \\
220521A & 5.60 & 1107466 & 3.70 & 13.6$\pm$2.7 & $-1.30_{-0.74}^{+0.89}$ &52.62$\pm$0.06 & 49$\pm$45 & Swift-BAT (CPL fit) \\
220117A & 4.96 & 1093592 & 3.50 & 51$\pm$2.9 & $-1.18_{-0.70}^{+0.84}$ & 52.86$\pm$0.06 & 53$\pm$33 & Swift-BAT (CPL fit) \\
220101A & 4.62 & 1091527 & 4.10 & 162$\pm$13 & $-1.27\pm0.03$ & 53.44$\pm$0.04 & 460$\pm$99 & Swift-BAT (PL fit) \\
211024B & 1.11 & 1081073 & 3.10 & 600$\pm$18 & $-1.61\pm0.09$ & 52.24$\pm$0.05 & 184$\pm$87 & Swift-BAT (PL fit) \\
210905A & 6.32 & 1071993 & 3.47 & 778$\pm$389 & $-0.97_{-0.63}^{+0.89}$ & 53.43$\pm$0.07 & 90$\pm$61 & Swift-BAT (CPL fit) \\
210822A & 1.74 & 1069788 & 4.12 & 186$\pm$47 & $-1.30\pm0.04$ & 52.88$\pm$0.03 & 419$\pm$107 & Swift-BAT (PL fit) \\
210731A & 1.25 & 1062336 & 3.43 & 21$\pm$2.8 & $-1.01\pm0.11$ & 51.65$\pm$0.07 & 1077$\pm$1022 & Swift-BAT (PL fit) \\
210722A & 1.15 & 1061223 & 3.57 & 50$\pm$10.6 & $-0.99_{-0.46}^{+0.51}$ & 51.93$\pm$0.04 & 95$\pm$106 & Swift-BAT (CPL fit) \\
210702A & 1.16 & 1058804 & 3.52 & 220$\pm$108 & $-1.53\pm0.09$ & 52.40$\pm$0.05 & 222$\pm$113 & Swift-BAT (PL fit) \\
210619B & 1.94 & 1056757 & 3.86 & 61$\pm$0.8 & $-1.40\pm0.02$ & 53.68$\pm$0.02 & 310$\pm$41 & Swift-BAT (PL fit) \\
210610B & 1.13 & 1054681 & 3.44 & 69$\pm$2.5 & $-1.13\pm0.03$ & 52.81$\pm$0.02 & 704$\pm$147 & Swift-BAT (PL fit) \\
210517A & 2.49 & 1048783 & 3.69 & 3.1$\pm$1.4 & $-0.98_{-0.98}^{+1.41}$ & 51.32$\pm$0.095 & 48$\pm$32 & Swift-BAT (CPL fit) \\
210504A & 2.08 & 1046782 & 2.93 & 143$\pm$9.3 & $-0.99_{-0.50}^{+0.57}$ & 52.45$\pm$0.043 & 76$\pm$57 & Swift-BAT (CPL fit) \\
210420B & 1.40 & 1044382 & 3.92 & 158$\pm$30 & $-1.47\pm0.19$ & 51.70$\pm$0.12 & 260$\pm$29 & Swift-BAT (PL fit) \\
210411C & 2.83 & 1042398 & 3.59 & 12.8$\pm$0.6 & $-1.62_{-0.45}^{+0.50}$ & 52.34$\pm$0.03 & 14.4$\pm$10.8  & Swift-BAT (CPL fit) \\
210210A & 0.72 & 1031728 & 3.74 & 6.6$\pm$0.6 & $-2.46\pm0.12$ & 51.27$\pm$0.05 & 37$\pm$15.8  & Swift-BAT (PL fit) \\
201104B & 1.95 & 1004168 & 3.72 & 8.7$\pm$0.2 & $-1.48\pm0.07$ & 52.01$\pm$0.05 & 257$\pm$105  & Swift-BAT (PL fit) \\
201024A & 1.0 & 1001514 & 3.28 & 5.0$\pm$2.2 & $-2.10\pm0.19$ & 51.37$\pm$0.09 & 67$\pm$51 & Swift-BAT (PL fit) \\
201021C & 1.07 & 1001130 & 4.05 & 35$\pm$2.2 & $-1.63\pm0.15$ & 51.38$\pm$0.08 & 178$\pm$142 & Swift-BAT (PL fit) \\
201020A & 2.90 & 1000926 & 3.15 & 14.4$\pm$4.8 & $-2.26\pm0.14$ & 52.41$\pm$0.13 & 51$\pm$28  & Swift-BAT (PL fit) \\
201014A & 4.56 & 1000255 & 3.50 & 36.2$\pm$8.1 & $-2.55_{-0.40}^{+0.34}$ & 52.50$\pm$0.42 & 32$\pm$41  & Swift-BAT (PL fit) \\
200205B & 1.47 & 954519 & 3.56 & 454$\pm$4.5 & $-1.35_{-0.30}^{ +0.32}$ & 52.59$\pm$0.022 & 61$\pm$16  & Swift-BAT (CPL fit) \\
191221B & 1.15 & 945521 & 3.55 & 48$\pm$16 & $-1.23\pm0.05$ & 52.58 $\pm$0.03 & 509$\pm$166 & Swift-BAT (PL fit) \\
191011A & 1.72 & 928924 & 3.46 & 7.4$\pm$0.9 & $-1.932_{-0.18}^{+0.17}$ & 51.39$\pm$0.12 & 92$\pm$72  & Swift-BAT (PL fit) \\
191004B & 3.50 & 927839 & 4.09 & 300$\pm$82 & $-1.10\pm0.15$ & 52.31$\pm$0.15 & 793$\pm$934  & Swift-BAT (PL fit) \\
190829A & 0.08 & 922968 & 4.75 & 57$\pm$47 & $-2.56_{-0.22}^{+0.21}$ &  49.97$\pm$0.05 & 32$\pm$23  & Swift-BAT (PL fit) \\
190719C & 2.47 & 915381 & 4.22 & 186$\pm$9.6 & $-1.46_{-0.27}^{+0.36}$ & 52.87$\pm$0.03 & 123$\pm$87  & Swift-BAT (CPL fit) \\
190114A & 3.38 & 883600 & 3.77 & 67$\pm$9.8 & $-2.06_{-0.23}^{+0.21}$ & 52.34$\pm$0.22 & 73$\pm$68  & Swift-BAT (PL fit) \\
190106A & 1.86 & 882252 & 3.51 & 78$\pm$2.1 & $-1.59\pm0.05$ & 52.53$\pm$0.04 & 194$\pm$56  & Swift-BAT (PL fit) \\
181110A & 1.51 & 871316 & 3.63 & 138$\pm$10.9 & $-1.91_{-0.22}^{+0.23}$ &  52.77$\pm$0.02 & 39$\pm$37  & Swift-BAT (CPL fit) \\
181020A & 2.94 & 867987 & 3.28 & 238$\pm$11.6 & $-1.25\pm0.06$ & 52.76$\pm$0.06 & 476$\pm$199  & Swift-BAT (PL fit) \\
181010A & 1.39 & 866434 & 4.02 & 15.6$\pm$2.2 & $-1.52\pm0.15$ & 51.37$\pm$0.09 & 232$\pm$197 & Swift-BAT (PL fit) \\
180728A & 0.12 & 850471 & 4.73 & 6.4$\pm$0.4 & $1.97\pm0.03$ & 50.98$\pm$0.01 & 86$\pm$9.9 & Swift-BAT (PL fit) \\
180720B & 0.65 & 848890 & 3.87 & 49$\pm$0.4 & $-1.17\pm0.01$ & 53.53$\pm$0.001 & 636$\pm$15 & Fermi-GBM (Band fit) \\
180624A & 2.86 & 844192 & 3.32 & 467$\pm$36 & $-1.88\pm0.1$ & 52.95$\pm$0.09 & 103$\pm$47 & Swift-BAT (PL fit) \\
180620B & 1.12 & 843211 & 3.64 & 47$\pm$1.3 & $-1.21\pm0.12$ & 52.49$\pm$0.005 & 176$\pm$50 & Fermi-GBM (Band fit) \\
180510B & 1.31 & 831816 & 3.83 & 134$\pm$53 & $-2.014_{-0.18}^{+0.17}$ & 51.99$\pm$0.11 & 79$\pm$59 & Swift-BAT (PL fit) \\
180404A & 1.00 & 821881 & 3.34 & 36$\pm$5.4 & $-1.95\pm0.12$ & 51.52$\pm$0.06 & 90$\pm$49 & Swift-BAT (PL fit) \\
180329B & 2.0 & 819490 & 3.58 & 214$\pm$49.4 & $-0.90_{-0.53}^{+0.60}$ & 52.51$\pm$0.04 & 50$\pm$9.7 & Swift-BAT (CPL fit) \\
180325A & 2.25 & 817564 & 3.69 & 93$\pm$1.5 & $-0.98_{-0.18}^{+0.19}$ & 52.90$\pm$0.013 & 258$\pm$372 & Swift-BAT (CPL fit) \\
180205A & 1.41 & 808625 & 3.40 & 15.4$\pm$1.5 & $-1.89$ & 50.56$\pm$0.02 & 101$\pm$11  & Fermi-GBM (PL fit) \\
171222A & 2.41 & 799669 & 3.31 & 80$\pm$4.6 & $-2.07$ & 50.49$\pm$0.01 & 71$\pm$7.7  & Fermi-GBM (PL fit) \\
171205A & 0.04 & 794972 & 4.05 & 191$\pm$34 & $-1.37\pm0.14$ & 49.02$\pm$0.04 & 340$\pm$294 & Swift-BAT (PL fit) \\
171020A & 1.87 & 780845 & 4.44 & 42$\pm$9.2 & $-1.03_{-0.20}^{+0.22}$ & 51.58$\pm$0.15 & 992$\pm$1697 & Swift-BAT (PL fit) \\
170714A & 0.79 & 762535 & 3.62 & 459$\pm$95 & $-1.50_{-0.34}^{+0.72}$ &  51.68$\pm$0.06 & 132$\pm$120 & Swift-BAT (CPL fit) \\
170705A & 2.01 & 760064 & 3.61 & 22.8$\pm$1.4 & $-0.99\pm0.07$ & 53.14$\pm$0.002 & 98$\pm$7.6 & Fermi-GBM (Band fit) \\
170607A & 0.56 & 756284 & 3.38 & 20.9$\pm$2.1 & $-1.40\pm0.04$ & 51.89$\pm$0.03 & 145$\pm$11.9 & Fermi-GBM (CPL fit) \\
\enddata
\end{deluxetable}

\begin{deluxetable}{ccccccccccccccc}
\tabletypesize{\scriptsize}
\renewcommand{\thetable}{}
\makeatletter
\renewcommand{\@makecaption}[2]{#2} 
\makeatother
\caption{\textbf{Table \ref{tab:prompt_parameters1}} (Continued)}
\label{tab:prompt_parameters2}
\tablehead{ 
\colhead{GRB} & \colhead{z} & \colhead{Trigger} & \colhead{$1 ~ \rm count$} & \colhead{$T_{90}$}  & \colhead{Spectral Index} & \colhead{$log(E_{\rm iso})$}  & \colhead{$E_{\rm pk}$} & \colhead{Instrument} \\
\colhead{} & \colhead{} & \colhead{} & \colhead{$\times10^{-11}$}  & \colhead{} & \colhead{} \\
 \colhead{}  & \colhead{} & \colhead{} &  \colhead{($\rm erg ~ cm^{-2}$)} & \colhead{(s)} & \colhead{} & \colhead{(erg)} & \colhead{(keV)}  &  \colhead{($\rm model ~type$)} 
}
\startdata
170604A & 1.33 & 755867 & 3.65 & 26.5$\pm$2.8 & $-1.30\pm0.12$ & 52.13$\pm$0.07 & 419$\pm$330 & Swift-BAT (PL fit) \\
170531B & 2.37 & 755354 & 3.38 & 170$\pm$9.1 & $-1.94\pm0.15$ & 52.39$\pm$0.12 & 92$\pm$61 & Swift-BAT (PL fit) \\
170519A & 0.82 & 753445 & 3.42 & 220$\pm$144 & $-1.95_{-0.24}^{+0.23}$ & 51.29$\pm$0.11 & 90$\pm$95  & Swift-BAT (PL fit) \\
170405A & 3.51 & 745797 & 3.61 & 79$\pm$0.6 & $-0.78\pm0.02$ & 54.29$\pm$0.001 & 267$\pm$9.3  & Fermi-GBM (Band fit) \\
170202A & 3.65 & 736407 & 3.21 & 38$\pm$11.7 & $-1.413_{-0.26}^{+0.27}$ & 52.97$\pm$0.02 & 113$\pm$1135 & Swift-BAT (CPL fit) \\
170113A & 1.99 & 732526 & 3.89 & 49$\pm$4.1 & $-1.96$ & 50.44$\pm$0.02 & 89$\pm$13.4  & Fermi-GBM (PL fit) \\
161219B & 0.15 & 727541 & 3.83 & 6.9$\pm$0.8 & $-1.84\pm0.08$ & 49.89$\pm$0.02 & 110$\pm$42 & Swift-BAT (PL fit) \\
161117A & 1.55 & 722604 & 3.45 & 122$\pm$0.7 & $0.87\pm0.03$ & 53.30$\pm$0.02 & 85$\pm$1.7  & Fermi-GBM (CPL fit) \\
161108A & 1.16 & 721234 & 3.49 & 116$\pm$11.7 & $-1.79\pm0.16$ & 51.53$\pm$0.09 & 125$\pm$98  & Swift-BAT (PL fit) \\
161017A & 2.01 & 718023 & 3.21 & 38$\pm$10.9 & $-1.09\pm0.09$ & 52.70$\pm$0.06 & 277$\pm$ 0.6  & Fermi-GBM (CPL fit) \\
161014A & 2.82 & 717500 & 3.75 & 37$\pm$1.5 & $-0.76\pm0.09$ & 53.05$\pm$0.03 & 170$\pm$12.8  & Fermi-GBM (CPL fit) \\
160804A & 0.74 & 707231 & 3.55 & 132$\pm$22 & $-1.09\pm0.06$ & 52.37$\pm$0.07 & 76$\pm$2.8 & Fermi-GBM (CPL fit) \\
160425A & 0.56 & 684098 & 3.11 & 305$\pm$15 & $-2.18_{-0.18}^{+0.17}$ & 51.25$\pm$0.07 & 59$\pm$42  & Swift-BAT (PL fit) \\
160410AS & 1.72 & 682269 & 4.68 & 96$\pm$51 & $-1.106_{-0.28}^{+0.29}$ & 51.56$\pm$0.20 & 766$\pm$1710 & Swift-BAT (PL fit) \\
160228A & 1.64 & 676595 & 4.40 & 99$\pm$24 & $-1.30\pm0.11$ & 51.84$\pm$0.08 & 414$\pm$301 & Swift-BAT (PL fit) \\
160227A & 2.38 & 676423 & 3.85 & 316$\pm$76 & $-1.55\pm0.10$ & 52.43$\pm$0.09 & 214$\pm$123 & Swift-BAT (PL fit) \\
160203A & 3.52 & 672525 & 3.73 & 17.4$\pm$2.3 & $-1.94_{-0.22}^{+0.21}$ & 52.39$\pm$0.22 & 91$\pm$87 & Swift-BAT (PL fit) \\
160131A & 0.97 & 672236 & 3.54 & 328$\pm$71 & $-1.21\pm0.05$ & 52.48$\pm$0.02 & 541$\pm$178 & Swift-BAT (PL fit) \\
160121A & 1.96 & 671231 & 4.14 & 10.5$\pm$2.4 & $-1.77\pm0.13$ & 51.67$\pm$0.10 & 128$\pm$82  & Swift-BAT (PL fit) \\
151215A & 2.59 & 667392 & 3.56 & 17.9$\pm$1.0 & $-1.99_{-0.34}^{+0.31}$ & 51.69$\pm$0.29 & 83$\pm$119  & Swift-BAT (PL fit) \\
151031A & 1.17 & 662330 & 3.27 & 5.0$\pm$0.3 & $-2.36_{-0.17}^{+0.16}$ & 51.22$\pm$0.09 & 43$\pm$27  & Swift-BAT (PL fit) \\
151029A & 1.42 & 662086 & 2.70 & 9.0$\pm$4.0 & $-0.28_{-1.03}^{+1.31}$ & 51.31$\pm$0.07 & 34$\pm$6.4  & Swift-BAT (CPL fit) \\
151027B & 4.06 & 661869 & 3.85 & 123$\pm$1.2 & $-0.95\pm0.29$ & 51.91$\pm$0.04 & 193$\pm$55  & Fermi-GBM (SBPL fit) \\
151027A & 0.81 & 661775 & 3.45 & 130$\pm$5.5 & $-1.54\pm0.06$ & 51.99$\pm$0.03 & 217$\pm$69  & Swift-BAT (PL fit) \\
151021A & 2.33 & 660671 & 3.50 & 7.2$\pm$0.6 & $-0.67\pm0.04$ & 53.21$\pm$0.01 & 269$\pm$10   & Fermi-GBM (CPL fit) \\
150821A & 0.76 & 652847 & 3.72 & 103$\pm$5.8 & $-1.24\pm0.02$ & 52.90$\pm$0.002 & 281$\pm$17  & Fermi-GBM (Band fit) \\
150818A & 0.28 & 652603 & 3.37 & 143$\pm$22 & $-1.88\pm0.09$ & 50.91$\pm$0.03 & 104$\pm$44  & Swift-BAT (PL fit) \\
150727A & 0.31 & 650530 & 2.78 & 49$\pm$4.0 & $-0.35\pm0.14$ & 51.04$\pm$0.06 & 209$\pm$17   & Fermi-GBM (CPL fit) \\
150323A & 0.59 & 635887 & 3.43 & 150$\pm$9.1 & $-1.74_{-0.08}^{+0.07}$ & 51.67$\pm$0.03 & 138$\pm$52 & Swift-BAT (PL fit) \\
150314A & 1.76 & 634795 & 3.92 & 10.7$\pm$0.1 & $-0.68\pm0.01$ & 53.82$\pm$0.001 & 347$\pm$7.9  & Fermi-GBM (Band fit) \\
150301B & 1.52 & 633180 & 3.71 & 17.1$\pm$4.6 & $-1.46\pm0.07$ & 51.87$\pm$0.23 & 270$\pm$120  & Swift-BAT (PL fit) \\
150206A & 2.09 & 630019 & 3.69 & 75$\pm$13 & $-1.34\pm0.05$ & 52.89$\pm$0.04 & 373$\pm$117 & Swift-BAT (PL fit) \\
141225A & 0.92 & 622476 & 3.90 & 56$\pm$4.9 & $-0.60\pm0.11$ & 51.93$\pm$0.04 & 258$\pm$27  & Fermi-GBM (CPL fit) \\
141221A & 1.45 & 622006 & 3.33 & 23.8$\pm$1.7 & $-1.18\pm0.11$ & 52.36$\pm$0.07 & 182$\pm$32 & Fermi-GBM (CPL fit) \\
141220A & 1.32 & 621915 & 3.68 & 7.6$\pm$0.9 & $-0.82\pm0.05$ & 52.40$\pm$0.03 & 178$\pm$9.1  & Fermi-GBM (CPL fit) \\
141121A & 1.47 & 619182 & 3.51 & 481$\pm$38 & $-1.43\pm0.48$ & 52.38$\pm$0.05 & 79$\pm$55 & Swift-BAT (CPL fit) \\
141004A & 0.57 & 614390 & 4.27 & 2.6$\pm$0.6 & $-1.78$ & 50.35$\pm$0.01 & 126$\pm$11.4  & Fermi-GBM (PL fit) \\
140907A & 1.21 & 611933 & 3.79 & 36$\pm$5.5 & $-1.01\pm0.06$ & 52.41$\pm$0.04 & 137$\pm$7.8  & Fermi-GBM (CPL fit) \\
140710A & 0.56 & 603954 & 3.93 & 3.0$\pm$2.2 & $-2.037_{-0.24}^{+0.22}$ & 50.28$\pm$0.09 & 76$\pm$74  & Swift-BAT (PL fit) \\
140703A & 3.14 & 603243 & 3.82 & 84$\pm$3.0 & $-1.28\pm0.05$ & 53.22$\pm$0.03 & 219$\pm$23  & Fermi-GBM (CPL fit) \\
140629A & 2.28 & 602884 & 3.27 & 38$\pm$11.6 & $-1.85\pm0.11$ & 52.39$\pm$0.09 & 109$\pm$55  & Swift-BAT (PL fit) \\
\enddata
~
\caption{\textbf{Note:} To obtain the overall uncertainty for $E_{pk}$ fitted with the cut-off power-law model to the Swift-BAT data, we sum the absolute values of the high and low errors. A few GRBs detected by Fermi-GBM have no reported errors for the spectral index in the Fermi-GBM catalog.}
\end{deluxetable}


\end{document}